\documentclass[%
reprint,
 amsmath,amssymb,
 aip,
 pop,
]{revtex4-2}

\usepackage{graphicx}
\usepackage{dcolumn}
\usepackage{bm}
\usepackage[colorlinks=true, linkcolor=blue, citecolor=blue]{hyperref}
\usepackage{xcolor}				

\usepackage{multirow}

\begin{document}

\preprint{APS/123-QED}

\title{Laser-Driven, Ion-Scale Magnetospheres in Laboratory Plasmas. II. Particle-in-cell Simulations}

\author{Filipe D. Cruz}
 \email{filipe.d.cruz@tecnico.ulisboa.pt}
 \affiliation{
 GoLP/Instituto de Plasmas e Fus\~{a}o Nuclear, \\ Instituto Superior T\'{e}cnico, Universidade de Lisboa, 1049-001 Lisboa, Portugal
}
\author{Derek B. Schaeffer}
\affiliation{Department of Astrophysical Sciences,  Princeton University, Princeton, NJ 08540, USA}
\author{F\'{a}bio Cruz}
\author{Luis O. Silva}
\email{luis.silva@tecnico.ulisboa.pt}
\affiliation{
 GoLP/Instituto de Plasmas e Fus\~{a}o Nuclear, \\ Instituto Superior T\'{e}cnico, Universidade de Lisboa, 1049-001 Lisboa, Portugal
}

\date{\today}

\begin{abstract}
Ion-scale magnetospheres have been observed around comets, weakly-magnetized asteroids, and localized regions on the Moon, and provide a unique environment to study kinetic-scale plasma physics, in particular in the collisionless regime. In this work, we present the results of particle-in-cell simulations that replicate recent experiments on the Large Plasma Device at the University of California, Los Angeles. Using high-repetition rate lasers, ion-scale magnetospheres were created to drive a plasma flow into a dipolar magnetic field embedded in a uniform background magnetic field. The simulations are employed to evolve idealized 2D configurations of the experiments, study highly-resolved, volumetric datasets and determine the magnetospheric structure, magnetopause location and kinetic-scale structures of the plasma current distribution. We show the formation of a magnetic cavity and a magnetic compression in the magnetospheric region, and two main current structures in the dayside of the magnetic obstacle: the diamagnetic current, supported by the driver plasma flow, and the current associated to the magnetopause, supported by both the background and driver plasmas with some time-dependence. From multiple parameter scans, we show a reflection of the magnetic compression, bounded by the length of the driver plasma, and a higher separation of the main current structures for lower dipolar magnetic moments.%
\end{abstract}

\maketitle


\section{Introduction}%

\par A vast range of space and astrophysical scenarios are driven by the rapid expansion of plasmas through space. Such examples include interplanetary coronal fast ejecta~\cite{Burlaga2001}, the expansion of the stellar material from supernova remnants~\cite{1990ApJ...356..549S}, and artificial magnetospheric releases of tracer ions~\cite{Krimigis1982}. When these expanding plasmas encounter obstacles of magnetic nature, the resultant interaction leads to highly nonlinear and complex dynamics. In the solar system, the interaction between the plasma flow (\textit{i.e.} the solar wind) and planetary-sized magnetic obstacles leads to the formation of magnetospheres~\cite{Russell1991}.%

\par The effective size of the magnetic obstacles is determined by the equilibrium position between the kinetic pressure of the solar wind and the magnetic pressure exerted by the planetary magnetic fields~\cite{Schield1969}. The region of equilibrium, called the magnetopause, can be described using the pressure balance derived from magnetohydrodynamics (MHD)%
\begin{equation}
    n_dm_{i,d}v_0^2 = \frac{B^2}{8\pi} \ ,
    \label{eq:pressure-equilibrium}
\end{equation}
where $n_d$ is the density of the solar wind, $v_0$ is its flow velocity, $m_{i,d}$ is the mass of its ions, and $B$ is the total magnetic field at the magnetopause. The total magnetic field can be written as $B = B_0 + B_\textrm{dip}$, where $B_0$ is the collective magnetic field and $B_\textrm{dip} = M / L_0^3$ is the magnetic field of the obstacle, often well described by a dipolar profile of magnetic moment $M$. The distance $L_0$ between the center of the dipole and the magnetopause, often referred to as the plasma standoff distance, measures the effective size of the magnetic obstacle.%

\par For planetary-sized magnetospheres, the obstacle size is typically tens of thousands of kilometers. However, magnetospheres with a few hundreds of kilometers are also observed in space environments such as the lunar surface. When the magnetic obstacle size is smaller or of the order of the ion kinetic scales of the plasma, \textit{i.e.} the ion skin depth or the ion gyroradius, the interaction with the solar wind results in ion-scale magnetospheres, or mini-magnetospheres.%

\par The study of mini-magnetospheres in past years was mainly motivated by the observation of crustal magnetic anomalies on the lunar surface~\cite{Lin1998,Halekas2008,Kato2010,Wieser2010,Kramer2021}. Although the Moon does not have a global magnetic field like Earth, it does have small localized regions of crustal magnetic field, of 10-100 nT over distances of 100-1000 km~\cite{Lin1998}, which are of the same order as the gyroradius of solar wind ions near the Moon's surface. As a result, when these regions of the lunar surface are exposed to the solar wind, mini-magnetospheres are formed. The deflection of charged particles off of lunar mini-magnetospheres commonly leads to the formation of ``lunar swirl'' structures~\cite{Bamford2012}. Similar interactions between the solar wind and small-sized patches of magnetic field also occur in other planets and natural satellites without a planetary magnetosphere, such as Mars~\cite{Lillis2013}, Mercury~\cite{doi:10.1126/science.1211001}, Ganymede~\cite{https://doi.org/10.1029/97GL02201}, and comets and asteroids~\cite{https://doi.org/10.1029/GL011i010p01022}.%

\par Multiple experiments have been performed in laboratory environments that replicate the interaction between plasma flows and magnetic obstacles. With a proper re-scaling of parameters~\cite{Ryutov_2002}, these experiments represent highly controlled configurations where a large variety of diagnostics can be used to obtain more accurate measurements than those obtained from the direct probing of astrophysical events. In experimental studies, fast-moving plasma flows are usually driven resorting to high-intensity lasers focused onto solid targets of plastic or metal composition~\cite{ablated,ablated2}. These laser-ablated plasmas can be mildly collisional or collisionless, replicating astrophysical conditions~\cite{Niemann2014a,Bondarenko2017}. By adding dipole field sources against the plasma flow, previous experiments of mini magnetospheres studied possible applications for spacecrafts~\cite{Winglee2007,Bamford2008,Bamford2014}, the formation of lunar swirls~\cite{Bamford2012}, and the conditions for the formation of magnetosphere features~\cite{Brady2009,Shaikhislamov2013, Shaikhislamov2014,Rigby2018}. Although these experiments achieved important breakthroughs in the study of ion-scale magnetospheric physics, they were limited to i) 1D measurements of the magnetic field and plasma density profiles and ii) fixed properties of the obstacle and plasma flow.%

\par Numerical simulations play a key role in interpreting and designing experiments. Early MHD simulations attempted to explain the formation and characteristics of lunar mini-magnetospheres and validate experimental and analytical models~\cite{Harnett2000,Harnett2002,Shaikhislamov2013}. Hybrid simulations were used to study the role of ion kinetic effects, and obtain conditions for the formation of magnetospheres~\cite{Blanco-Cano2004} and replicate previous experimental results~\cite{Gargate2008}. However, these simulations do not resolve the electron scales and do not capture important kinetic effects on the magnetosphere's boundary, \textit{e.g.} charge separation effects and nonthermal particle distributions. Particle-in-cell (PIC) simulations were used to fully resolve the micro-physics of these systems and study its role in the formation of lunar mini-magnetospheres~\cite{Kallio2012,Deca2014,Deca2015,Zimmerman2015,Bamford2016}, the scaling of their properties with solar wind speed and magnetic field orientation~\cite{Deca2021} and the conditions for the formation of collisionless shocks~\cite{Cruz2017}.%

\par In this work, we use PIC simulations of ion-scale magnetospheres to interpret the results of recent experiments~\cite{Schaeffer2021} performed at the LArge Plasma Device (LAPD), University of California, Los Angeles. In these experiments, fast collisionless plasma flows generated by high-repetition-rate lasers were collided with the magnetized ambient plasma provided by the LAPD and with a dipolar magnetic field obstacle, leading to the formation of ion-scale magnetospheres. Using motorized probes, high spatial and temporal resolution measurements of the magnetic field allowed characterization of 2D magnetic field and current density structures. Apart from validating the experimental results, the simulations presented in this work explore a set of upstream and magnetic parameter scans and configurations not accessible in the laboratory to determine the importance of each system parameter on the magnetospheric properties. The simulations show that the background ions, and then the driver ions, are responsible for the formation of the magnetopause observed in the experiments. They also show that a reflection of the downstream magnetic compression is observed for certain parameters of the driver plasma, and that the distance between the main current features is dependent on the dipolar and driver plasma parameters. 

\par This paper is organized as follows. In Sec.~\ref{sec:experiments}, we briefly review the LAPD experiments and their main results. In Sec.~\ref{sec:simulations}, we present PIC simulations of ion-scale magnetospheres. In Sec.~\ref{sec:numerical-methods}, we outline the standard configuration and parameters used for the simulations. In Sec.~\ref{sec:additional}, we provide an overview of the temporal evolution of these systems and show that the simulations agree with the results of the LAPD experiments. We discuss the origin of the structures observed in current density and magnetic field synthetic diagnostics and use particle phase spaces to interpret them. In Sec.~\ref{sec:length}, we present the results for different lengths of the plasma flow and define the conditions required to reproduce the features observed experimentally. The coupling between the laser-ablated driver and background plasmas is characterized in Sec.~\ref{sec:density} with simulations with different driver densities. In Sec.~\ref{sec:momentum}, different magnetic moments are considered, and we show that the main current density features are highlighted and more easily visible for weaker magnetic obstacles.
In Secs.~\ref{sec:realistic} and~\ref{sec:finite}, we discuss and illustrate the validity of the key simplifications and approximations used for the parameter scans presented in Secs.~\ref{sec:additional}-\ref{sec:momentum}. Finally, we outline the conclusions of this work in Sec.~\ref{sec:conclusions}.%

\par This paper is the second part of a two part series. Detailed experimental results are presented in Part I~\cite{Schaeffer2021}. 

\section{LAPD Experiment}
\label{sec:experiments}

\par A new experimental platform has been developed on the LAPD to study mini-magnetospheres. The platform combines the large-scale magnetized ambient plasma generated by the LAPD, a fast laser-driven plasma, and a pulsed dipole magnet, all operating at high-repetition-rate ($\sim1$ Hz). In the experiments, a supersonic plasma is ablated from a plastic target and then expands into the dipole magnetic field embedded in the ambient magnetized plasma. By measuring 2D planes of the magnetic field over thousands of shots, detailed maps of the magnetic field evolution are constructed.  Additional details on the platform and results can be found in Part I.

\par
\begin{figure}[ht]
    \includegraphics[width=0.85\linewidth]{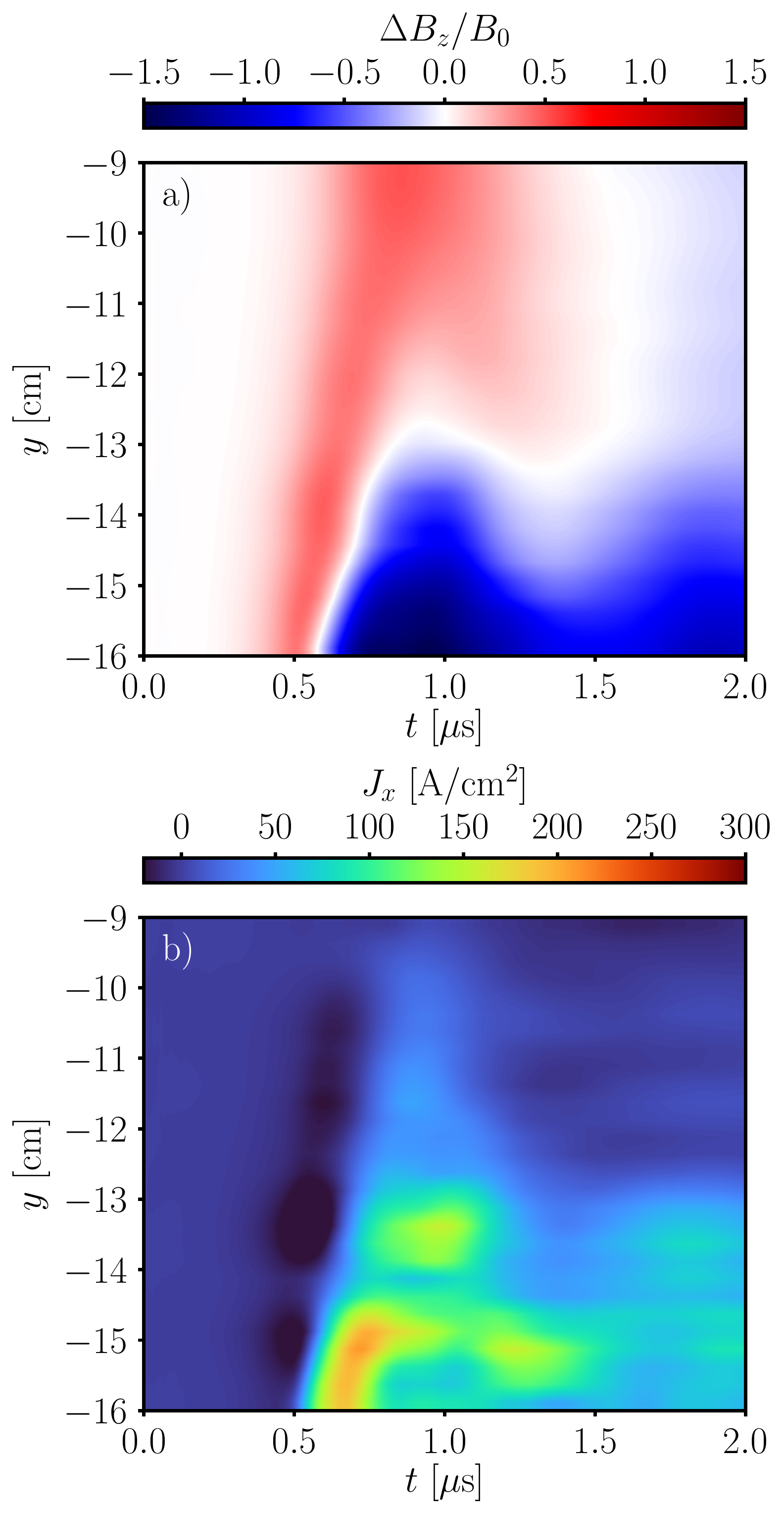}
    \caption{\label{fig:experiment} LAPD experimental results for the temporal evolution of a) the variation of the magnetic field $\Delta B_z$ and b) the current density $J_x$ at $x=z=0$. The experimental results are discussed with more detail in Part I.}
\end{figure}

 Example results are shown in Fig.~\ref{fig:experiment} for the measured change in magnetic field $\Delta B_z = B_{z}-B_{z,\textrm{initial}}$ and the current density $J_x = \partial \Delta B_z/\partial y$.  Here, $B_{z}$ is the total magnetic field, $B_{z,\textrm{initial}}=B_0 + B_\textrm{dip}$ is the total initial magnetic field, $B_0$ is the background LAPD field, and $B_\textrm{dip}$ is the dipole magnetic field.  These results are taken along $y$ at $x=0$ from the $z=0$ plane probed experimentally. In the experiments, the dipole is centered at $(x,y,z)=(0,0,0)$ and has a magnetic moment $M=475$ Am$^2$.

\par As seen in Fig.~\ref{fig:experiment}(a), the expanding laser-driven plasma creates a leading magnetic field compression followed by a magnetic cavity.  The cavity reaches a peak position of $y\approx-13$ cm, while the compression propagates closer to the dipole before being reflected back towards the target.  The current density in Fig.~\ref{fig:experiment}(b) shows two prominent structures.  Following the expansion of the magnetic cavity is a diamagnetic current, which reaches a peak position of $y\approx-15$ cm before stagnating for approximately 1 $\mu$s and then dissipating.  Ahead of the diamagnetic current is the magnetopause current near $y\approx-13.5$ cm, which lasts for about 0.5 $\mu$s.

\par Here, we aim to qualitatively model these experiments in order to address key questions that aid in the interpretation of the experimental results. In particular, simulations can explain the role of each system component in the features observed, address which plasma component (ambient or laser-driven) is responsible for the features observed and which pressure balances are most relevant. 

\par For convenience, the notations used in this paper are different for the ones used in Part I. Here, we use the CGS system, and the axis system is rotated from the used in Part I.

\section{PIC simulations}
\label{sec:simulations}

\subsection{Configuration of the simulations} \label{sec:numerical-methods}

\par Motivated by the results of experiments described in Sec.~\ref{sec:experiments}, we performed 2D simulations with OSIRIS, a massively parallel and fully relativistic PIC code~\cite{Fonseca2002,Fonseca2013}. With PIC simulations, we can accurately resolve the plasma kinetic scales characteristic of mini-magnetospheres dynamics.%

\par
\begin{figure}[ht]
    \includegraphics[width=0.85\columnwidth]{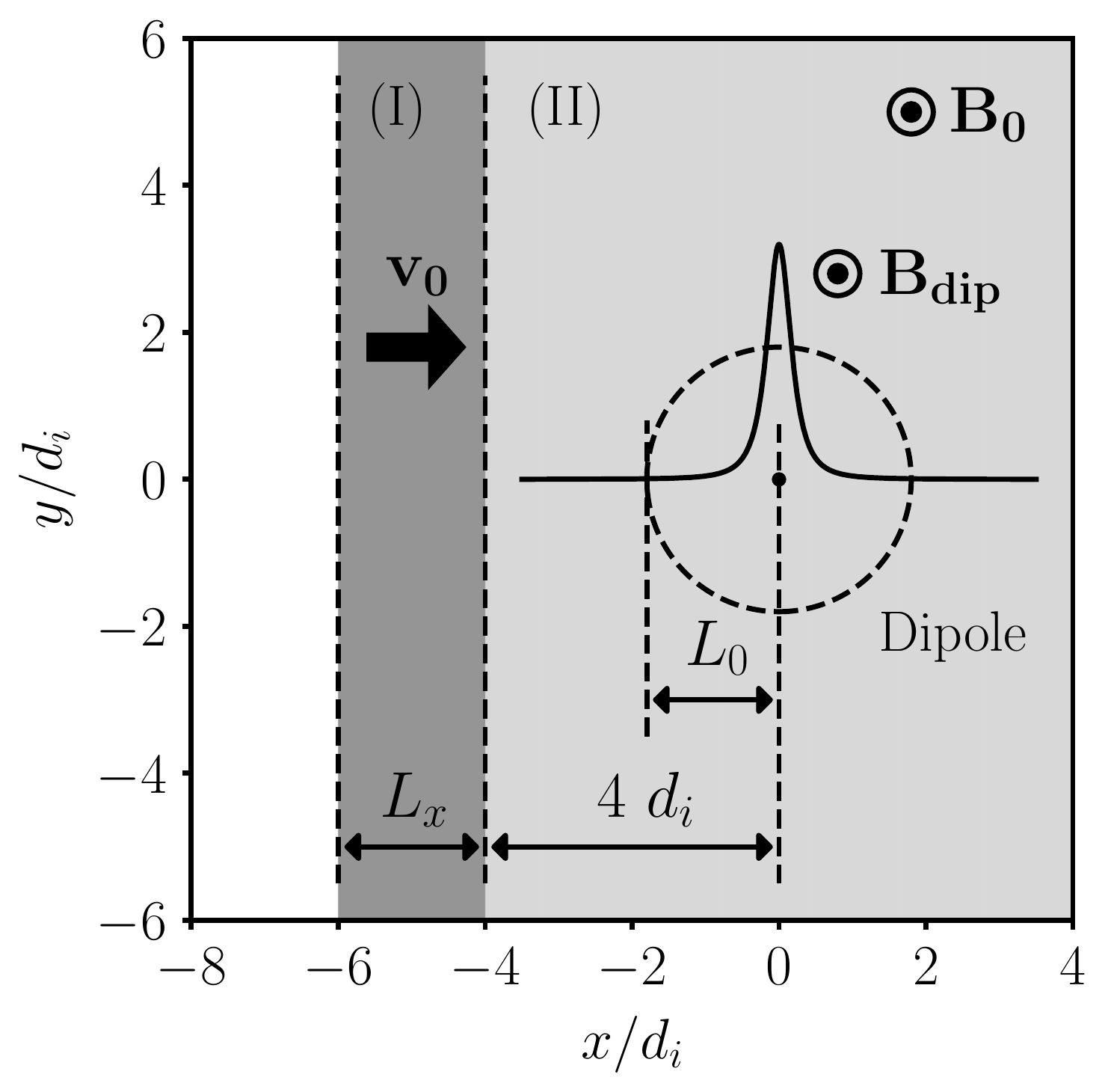}
    \caption{\label{fig:config} Schematic illustration of the initial setup of the 2D PIC simulations performed. The system considers a vacuum region at the left, a driver plasma (I) of density $n_d$ and length $L_x$, travelling to the right with flow velocity $v_0$, and a background plasma (II) with constant density $n_0$ and with an internal magnetic field $B_0$. A dipole is included at the center of the background region. Both the uniform and the dipolar magnetic fields are oriented in the $z$ direction. An illustration of the effective magnetic obstacle created by the dipole and of the magnetic field profile at $y=0$ are also shown in a dashed circumference and in a solid black line, respectively.}%
\end{figure}

 The numerical simulations presented in this work stem from a simplified description of the LAPD experimental setup, represented in Fig.~\ref{fig:config}. In these simulations, a driver plasma moves against a background plasma permeated by a uniform magnetic field $\mathbf{B_0}$ and a dipolar magnetic field $\mathbf{B_{dip}}$. $\mathbf{B_0}$ and $\mathbf{B_{dip}}$ are oriented along the $z$ direction and are transverse to the driver plasma flow. Since the most relevant dynamics of the simulations occurs at the ion kinetic scales, all the spatial scales are normalized to the ion skin depth of the background plasma $d_i=c/\omega_{pi}=\sqrt{m_{i,0}c^2/4\pi n_0e^2}$, where $c$ is the speed of light in vacuum, $\omega_{pi}$ is the ion plasma frequency, $m_{i,0}$ is the mass of the background plasma ions, $n_0$ is the background density, and $e$ is the electron charge. In turn, the temporal scales are normalized to $1/\omega_{ci}$, where $\omega_{ci} = eB_0/m_{i,0}c$ is the ion cyclotron frequency of the background. The simulation box is a 12 $d_i$ $\times$ 12 $d_i$ area with open and periodic boundary conditions in the $x$ and $y$ directions, respectively. The flow is in the $x$ direction and the size of the simulation domain in the $y$ direction is large enough to avoid re-circulation of the particles through the whole interaction. The simulations considered 25 particles per cell per species. To resolve the dynamics of the electron kinetic scales, we used 10 grid cells per electron skin depth $d_e=d_i\sqrt{m_e/m_{i,0}}$ in both $x$ and $y$ directions, where $m_e$ is the electron mass.

\par The driver plasma, shown in region I in Fig.~\ref{fig:config}, represents ideally the plasma ablated from the plastic target in the experiments. We assume that this driver has a length $L_x$ that is typically 2 $d_i$, and a width $L_y$ that is typically infinite. It has a constant density $n_{d}$, and it is initialized moving to the right side with initial flow velocity $v_0$. The driver is composed of an electron species and a single ion species, with ion mass $m_{i,d}$. Because the driver plasma is reflected during the interaction with the background, an empty region at the left of the driver was added to accommodate the reflecting particles.%

\par The background plasma is represented in region II. It is an 8 $d_i$ length and infinite width plasma and it has uniform density $n_0$. The initial interface between the driver and background plasma is located at $x_B=-4\ d_i$. Like the driver plasma, it has an electron species and a single ion species, of mass $m_{i,0}$. The background plasma is magnetized with an internal uniform magnetic field $\mathbf{B_0} = B_0 \mathbf{\hat{z}}$, and its magnitude is defined such that the Alfvénic Mach number of the flow, $M_A \equiv v_0/v_A = v_0\sqrt{4\pi n_0m_{i,0}}/B_0$ matches the peak experimental value $M_A=1.5$, where $v_A$ is the Alfvén velocity.%

\par A dipolar magnetic field is externally imposed in our simulations (\textit{i.e.}, it is added to the plasma self-consistent electromagnetic fields to advance particle momenta but is not included in Maxwell's equations to advance the fields). The dipole is centered at $(x,y)=(0,0)$ and its associated magnetic field is $\mathbf{B_{dip}} = B_\mathrm{dip} \mathbf{\hat{z}}$, with $B_\mathrm{dip} = M / r^3$, where $M$ is the dipolar magnetic moment, $r = \sqrt{x^2 + y^2+\delta^2}$ is the distance to the origin of the dipole and $\delta=0.25\ d_i$ is a regularization parameter. For most simulations, the magnetic moment $M$ was chosen such that the expected standoff, obtained from Eq.~\eqref{eq:pressure-equilibrium}, is similar to the experimental value $L_0=1.8\ d_i$. For this particular magnetic moment, the total initial magnetic field $B_0+B_\mathrm{dip}$ is $\approx 3.0~B_0$ at the standoff distance. Near the interface between the driver and background plasmas, the magnetic field of the dipole is relatively small and the initial magnetic field is $\approx 1.2~B_0$.%

\par In this work, we present simulations with different drivers and magnetic dipole moments. All the simulations presented here, and their respective parameter sets, are listed in Table~\ref{tab:runs}. Simulations B-G are discussed through Sec.~\ref{sec:simulations} on equally labeled subsections. Simulation B is used to discuss the overall dynamics of the system, while simulations C, D, and E illustrate the role of the driver length, the density ratio, and the magnetic moment, respectively. Simulations F show the results for more realistic choices of parameters and simulation G for a more realistic driver shape. The physical parameters of the simulations (\textit{e.g.} $M_A$, $L_0/d_i$) were adjusted to be similar to the LAPD experiments, whereas other parameters (\textit{e.g.} $m_i/m_e$, $v_0$, $v_{the}$) were chosen to make simulations computationally feasible. The experimental and numerical parameters are presented in Table~\ref{tab:parameters} and compared with lunar mini-magnetospheres.%

\begin{table*}
    \caption{\label{tab:runs} List of simulations performed and their parameters. $v_{the,x}$ and $v_{thi,x}$ represent the $x$ component of the electron and ion thermal velocities, respectively. All the runs considered $v_{th,x}=v_{th,y}=v_{th,z}$ for the electrons and ions.}%
    \begin{ruledtabular}
        \begin{tabular}{cccccccccc}
            \textrm{Name} & $v_{the,x}/v_0$ & $v_{thi,x}/v_0$ & $n_d/n_0$ & $m_{i}/m_e$ & $m_{i,0}/m_e$ & $L_x/d_i$ & $L_y/d_i$ & $L_0/d_i$\\ \colrule
            \textrm{B/D2/E2} & 0.1 & 0.01 & 2 & 100 & 100 & 2 & $+\infty$ & 1.8 \\
            \textrm{C1} & 0.1 & 0.01 & 2 & 100 & 100 & 1 & $+\infty$ & 1.8 \\
            \textrm{C2} & 0.1 & 0.01 & 2 & 100 & 100 & 4 & $+\infty$ & 1.8 \\
            \textrm{C3} & 0.1 & 0.01 & 2 & 100 & 100 & $+\infty$ & $+\infty$ & 1.8 \\
            \textrm{D1} & 0.1 & 0.01 & 1 & 100 & 100 & 2 & $+\infty$ & 1.8 \\
            \textrm{D3} & 0.1 & 0.01 & 4 & 100 & 100 & 2 & $+\infty$ & 1.8 \\
            \textrm{E1} & 0.1 & 0.01 & 2 & 100 & 100 & 2 & $+\infty$ & 2.3 \\ 
            \textrm{E3} & 0.1 & 0.01 & 2 & 100 & 100 & 2 & $+\infty$ & 1.4 \\ \colrule
            \textrm{F1} & 0.1 & 0.002 & 2 & 1836 & 1836 & 2 & $+\infty$ & 1.8 \\
            \textrm{F2} & 2.5 & 0.033 & 2 & 1836 & 1836 & 2 & $+\infty$ & 1.8 \\
            \textrm{F3} & 2.5 & 0.033 & 2 & 100 & 100 & 2 & $+\infty$ & 1.8 \\
            \textrm{G} & 0.1 & 0.01 & 2 & 100 & 100 & 2 & 6 & 1.8 \\
        \end{tabular}
    \end{ruledtabular}
\end{table*}

\par In most simulations, we considered a reduced mass ratio $m_i / m_e = 100$, a flow velocity $v_0 / c = 0.1$, and cold plasmas to reduce the required computational resources, allow extended scans over the different parameters of the system, and simplify our analysis. The thermal effects are negligible for the main results, and the chosen ion-to-electron mass ratio is high enough to ensure sufficient separation between electron and ion spatial and temporal scales. We confirm the validity of our assumptions in Sec.\ref{sec:realistic}.%

\par In most of the simulations presented in this work, we have assumed that ions and electrons are initially in thermal equilibrium, and thus used the electron thermal velocities $v_{the}$ shown in Table~\ref{tab:runs}, to compute the ion thermal velocities $v_{thi}$. Because we aim to study the role of the hydrogen ions of the experimental driver in the interaction with the background plasma, these simulations considered equal ion masses for the driver and background plasmas, \textit{i.e.} $m_{i,d}=m_{i,0}$.%

\begin{table*}
    \caption{\label{tab:parameters} Typical parameters associated with lunar mini-magnetospheres~\cite{Russell1991,Bamford2012,Lin1998}, the range of parameters of LAPD~\cite{Schaeffer2021} and the canonical simulation B. The parameters are written in both physical and normalized units to facilitate the comparison between the space, the laboratory environments and the PIC simulations. The experimental parameters are presented in ranges of values computed with the possible LAPD values for the flow velocity $v_0$, the density $n_0$ and the temperature $T$. The plasma parameters shown for lunar mini-magnetospheres are relative to the solar wind, while for the experiments and the simulations, they are relative to the background plasma. The ion data shown corresponds to Hydrogen ions. The magnetic field $B_{\textrm{std}}$ is calculated at the standoff position, \textit{i.e.}, at a distance $L_0$ from the center of the obstacle.}%
    \begin{ruledtabular}
        \begin{tabular}{cccccc}
            \multirow{2}{*}{\textrm{Parameters}} & \multicolumn{2}{c}{Lunar mini-magnetospheres} & \multicolumn{2}{c}{LAPD experiments} & PIC simulations \\ 
            & Physical units & Normalized units & Physical units & Normalized units & Normalized units\\ \colrule \\ [-9 pt]
            Flow velocity, $v_0$ & $400$ km/s & $10^{-3}$ $c$ & 200-300 km/s & 0.7-1.0$\times10^{-3}$ $c$ & 0.1 $c$ \\
            Density, $n_0$ & 5 cm$^{-3}$ & --- & $10^{12}$-$10^{13}$ cm$^{-3}$ & --- & ---\\
            Mass ratio, $m_i/m_e$ & --- & 1836 & --- & 1836 & 100 \\
            Ion skin depth, $d_i$ & 100 km & --- & 7-23 cm & --- & --- \\
            Electron skin depth, $d_e$ & 2 km & 2$\times10^{-2}$ $d_i$ & 0.2-0.5 cm & 0.7-7.0$\times10^{-2}$ $d_i$ & 0.1 $d_i$ \\
            Magnetic obstacle size, $L_0$ & 300 km & 3 $d_i$ & 14-18 cm & 0.6-2.5 $d_i$ & 1.8 $d_i$\\
            Internal magnetic field, $B_{0}$ & $10^{-4}$ G & $10^{-2}$ $m_ec^2/ed_e$ & 300 G & 3-9$\times10^{-2}$ $m_ec^2/ed_e$ & 0.67 $m_ec^2/ed_e$ \\
            Ion gyroradius, $\rho_i$ & 500 km & 5 $d_i$ & 7-10 cm & 0.3-1.5 $d_i$ & 1.5 $d_i$ \\
            Electron gyroradius, $\rho_e$ & 800 m & $8\times10^{-3}$ $d_i$ & 4-6$\times10^{-3}$ cm & 2-8$\times10^{-4}$ $d_i$ & 0.15 $d_i$ \\
            Ion gyroperiod, $\omega_{ci}^{-1}$ & 1 s & --- & 230-520 ns & --- & --- \\
            Alfvén velocity, $v_A$ & 80 km/s & $3\times10^{-4}$ $c$ & 140-980 km/s & 0.5-3.3$\times10^{-3}$ $c$ & 0.067 $c$ \\
            Alfvénic Mach number, $M_A$ & --- & 5 & --- & 0.3-1.5 & 1.5 \\
            Temperature, $T$ & 5 eV & --- & 1-10 eV & --- & --- \\
            Electron thermal velocity, $v_{the}$ & 1500 km/s & 4 $v_0$ & 730-2300 km/s & 2.4-11.5 $v_0$ & 0.1 $v_0$ \\
            Ram pressure, $n_0m_iv_0^2$ & 1 nPa & --- & 70-1500 Pa & --- & --- \\
            Standoff magnetic field, $B_{\textrm{std}}$ & $5\times10^{-4}$ G & 0.07 $m_ec^2/ed_e$ & 100-600 G & 0.02-0.2 $m_ec^2/ed_e$ & 2.0 $m_ec^2/ed_e$ \\
        \end{tabular}
    \end{ruledtabular}
\end{table*}

\subsection{\label{sec:additional} Evolution and main features of the system}

\par
\begin{figure*}[t]
    \includegraphics[width=0.98\linewidth]{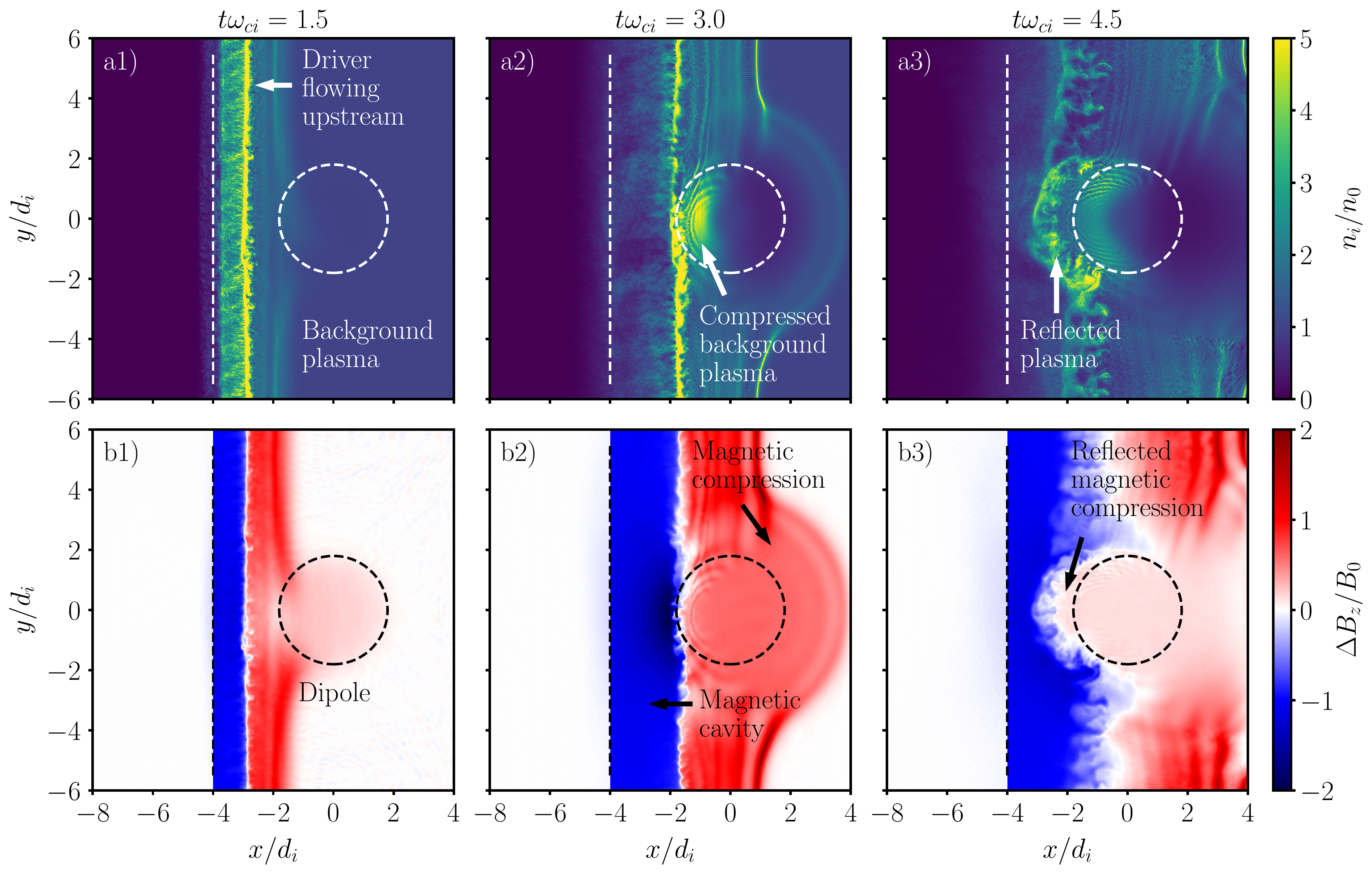}
    \caption{\label{fig:movie} Spatiotemporal evolution of a) the total ion density and b) the variation of the $z$ component of the magnetic field in simulation \textrm{B} (see Table~\ref{tab:runs} for a list of parameters). Columns 1-3 correspond to three different times in the simulation. The vertical and circular dashed lines mark the initial border between the driver and background plasma and the dipolar magnetic obstacle with radius $L_0$, respectively.}%
\end{figure*}

 To identify the main magnetospheric and kinetic-scale structures that arise from the initial configuration, simulation \textrm{B} was performed. It considered a driver with length $L_x=2\ d_i$ and density $n_d=2\ n_0$ (twice the background density). Figs.~\ref{fig:movie} a1-3) represent the total ion density $n_i=n_{i,d}+n_{i,0}$, for three different times, and Figs.~\ref{fig:movie} b1-3) show the variation of the $z$ component of the magnetic field, from its initial value, $\Delta B_z=B_z-B_{z,\textrm{initial}}$.%

\par In Fig.~\ref{fig:movie} a1), we see the total ion density for an early time ($t\omega_{ci}=1.5$). Given the small distance propagated by the driver plasma at this time, the dipolar magnetic field does not significantly affect the interaction between the plasmas. For this reason, we can express the early system as a driver flowing against a uniform magnetized background plasma. 
In Fig.~\ref{fig:movie} b1), we observe that this interaction creates a region of compressed magnetic field in the downstream region, where the background plasma is located, and expels the magnetic field in the region of the driver, leading to a magnetic cavity in the upstream region with approximately null magnetic field~\cite{Bondarenko2017}.%

\par In Figs.~\ref{fig:movie} a2) and b2), we start to observe the effects of the dipolar magnetic field for a later time ($t\omega_{ci}=3.0$). As the magnetic pressure exerted against the plasmas increases, a region of compressed background plasma forms in front of the dipole, as Fig.~\ref{fig:movie} a2) shows. After the interaction between the background and the dipole, the magnetic field pressure becomes large enough to counterbalance the kinetic pressure of the driver, reflecting it upstream. This can be seen seen in Fig.~\ref{fig:movie} a3) for a subsequent time ($t\omega_{ci}=4.5$). After the reflection, there is no longer a plasma flow pushing the magnetic compression forward or holding the decompression by the left side of the background region, and as a result, the region near the dipole quickly decompresses --- see Fig.~\ref{fig:movie} b3).%

\par
\begin{figure}[ht]
    \includegraphics[width=0.85\columnwidth]{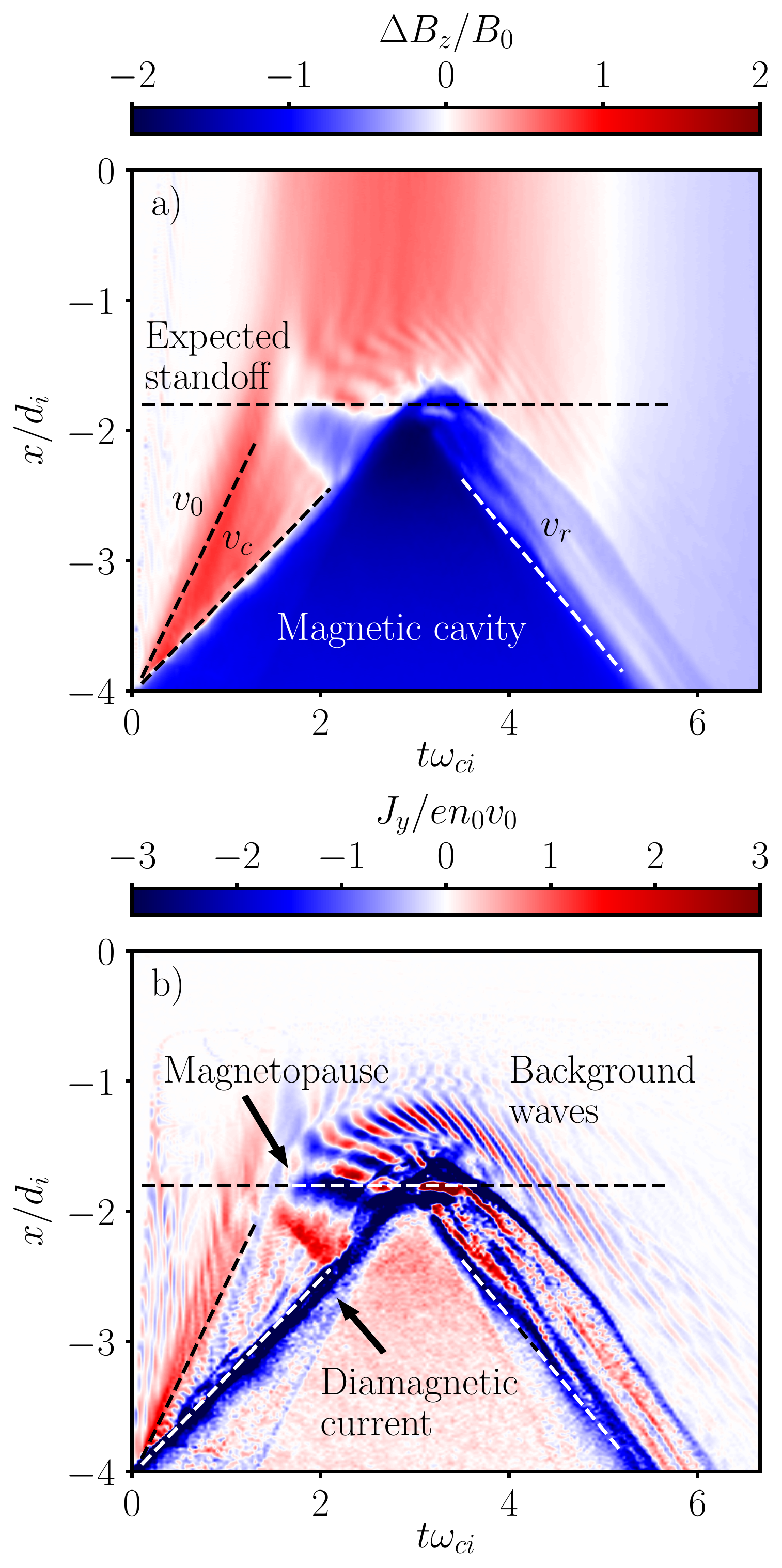}
    \caption{\label{fig:standard} Temporal evolution of a) the variation of the magnetic field $B_z$ and b) current density $J_y$ at $y=0$ for the simulation \textrm{B}. The driver has a 2 $d_i$ length and a density $n_d = 2\ n_0$. The dashed lines have slopes that match the flow velocity $v_0$, the coupling velocity $v_c$ and the reflection velocity $v_r$.}%
\end{figure}

 To compare the numerical results with the experimental data shown in Fig.~\ref{fig:experiment}, synthetic diagnostics were obtained from the simulations. In Fig.~\ref{fig:standard}, the variation of the magnetic field $\Delta B_z$ and the density current $J_y$ measured at the axis of symmetry $y=0$ and as a function of time are plotted for simulation \textrm{B}. These diagnostics are important to comprehend the system dynamics, due to the importance of the $z$ direction of the magnetic field in the motion of the particles.%

\par The main features of Fig.~\ref{fig:standard} are consistent with the experimental results. In the magnetic field plot of Fig.~\ref{fig:standard} a), both the upstream magnetic cavity and the downstream magnetic compression are present. Between $t\omega_{ci}=0$ and $t\omega_{ci}\approx1.5$, the system behaves approximately as a driver piston moving against a uniform magnetized plasma. As the driver pushes the background plasma and magnetic field, the discontinuity that separates these two media travels at constant coupling velocity $v_c < v_0$, measured as $v_c\approx0.49\ v_0$ for this simulation. The leading edge of the compression of the magnetic field travels with a velocity close to $v_0$ for the runs considered.%

\par The driver experiences increasingly higher magnetic fields until the magnetic pressure is enough to reflect the driver near the expected standoff $x_0 = -L_0$, at $t\omega_{ci}\approx3$. The magnetic cavity and magnetic compression are also reflected, and the boundary between these two regions travels with a velocity $v_r$ after reflection. The background magnetic decompression is seen after $t\omega_{ci}=5$.%

\par In the current density plot of Fig.~\ref{fig:standard} b), we can observe the diamagnetic current that supports the magnetic field gradient between the driver and background plasmas and that identifies the leading edge of the magnetic cavity. During the driver reflection, this current branches into multiple components due to the multi-stream velocity distributions developed in the driver and background plasmas. We can also verify that this structure is reflected near the expected standoff $x_0 =-L_0$. Between $t\omega_{ci}\approx2$ and $t\omega_{ci}\approx3$, a second current structure is present in the background region. It is associated with the magnetopause of the system and the small decompressed field region that we see in Fig.~\ref{fig:standard} a), and it arises from the interaction of the accelerated background ions with the dipole, as we show in Sec.~\ref{sec:momentum}. The presence of these two current structures is consistent with the experimental results.%

\par In Fig.~\ref{fig:standard} b) we can also see the formation of waves in the background plasma, near the dipole region. These waves are excited in regions of highly non-uniform density and magnetic field, and have periods and wavelengths between the ion and electron kinetic scales. We have verified that their properties change significantly for different ion thermal velocities. In particular, we have found these waves to be more clearly excited for lower ion temperatures, which may explain why these waves have not been observed in the experiments performed at the LAPD. A detailed characterization of these waves and the conditions for their formation is out of the scope of this paper, and shall be addressed in a future work.%

\par
\begin{figure*}[t]
    \includegraphics[width=0.95\linewidth]{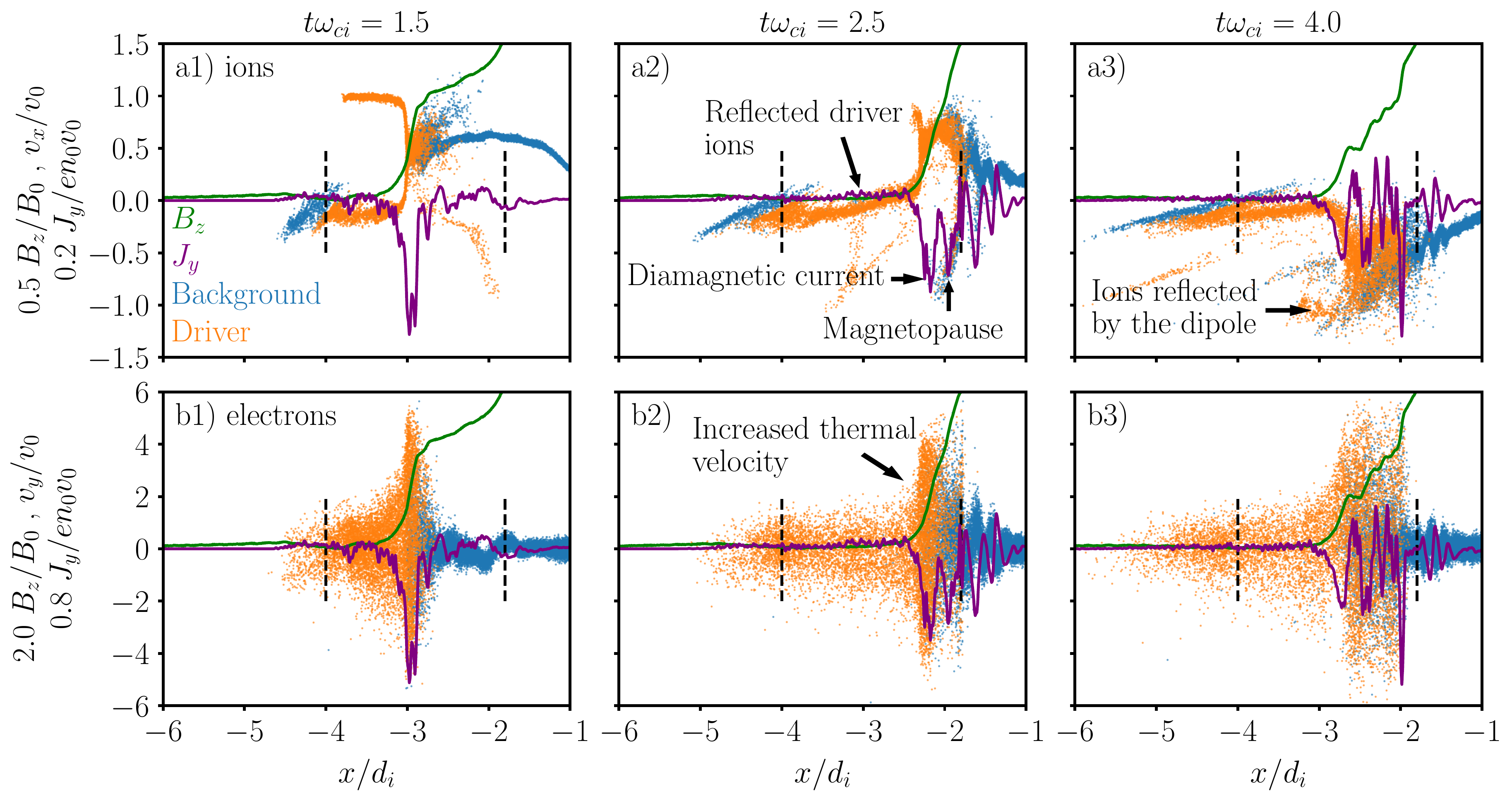}
    \caption{\label{fig:phase-space} Ion (a) and electron (b) phase spaces, and magnetic field $B_z$ and current density $J_y$ profiles at $y=0$, for simulation \textrm{B} and for three different times (1-3). The particles shown were randomly selected in the region $-0.2\ d_i < y < 0.2\ d_i$. The frames labeled a1) to a3) show the $v_x$ velocity of the ions, while the frames labeled b1) to b3) show the $v_y$ velocity of the electrons. Blue/orange markers correspond to background/driver plasma particles. The green and purple lines correspond to the magnetic field $B_z$ and current density $J_y$, respectively. The left dashed line marks the initial border between the driver and the background plasmas, and the right dashed line marks the expected standoff $x_0=-L_0$.}%
\end{figure*}

 To better understand the particle motion during the events described, we show in Fig.~\ref{fig:phase-space} the phase spaces of ions and electrons located near $y=0$. For the ions, the $x$ component of the velocity of the particles is presented, to illustrate their reflection and accumulation, while for the electrons, the $y$ component is shown instead, to show the formation of the currents. The magnetic field $B_z$ and the current density $J_y$ profiles for $y=0$ are also represented. Once again, we used the parameter set \textrm{B} of Table~\ref{tab:runs}.%

\par Fig.~\ref{fig:phase-space} a1) shows the $v_x$ velocity of the ions when the dipole field is still negligible. The ions initially move upstream with velocity $v_0$ until they interact with the background field. After reaching the background, they are mostly decelerated and reflected by electric field in the interface between the plasmas~\cite{Bondarenko2017}, and end up with a flow velocity that is close to zero for the simulation considered. The reflection occurs near the boundary of the magnetic cavity, which moves with velocity $v_c$ through the background, as mentioned above. During this stage, the background ions accelerate from rest to velocities of average close to $v_c$.%

\par The driver and the accelerated background ions continue to approach the dipole until they are reflected. This can be seen in Fig.~\ref{fig:phase-space} a2). During this interaction, two main current structures are visible in the $J_y$ profile. The first one (from the left) corresponds to the typical diamagnetic current, while the second one corresponds to the magnetopause. To the right of these two main current structures, we can see the background waves observed in Fig.~\ref{fig:standard} b). In Fig.~\ref{fig:phase-space} a3), the driver ions are totally reflected. The ions reflected by the dipole obtain a velocity close to $-v_0$, while the magnetic cavity moves back with velocity $v_r$.%

\par Because the simulation considers a cold plasma approximation, the ion thermal velocities remain small most of the time, except for the boundary between the two plasmas, where the velocity of the ions changes abruptly. The same does not occur for the electrons. We can see in the $v_y$ velocity of the electrons, represented in Figs.~\ref{fig:phase-space} b1) to b3) that, although the electron thermal velocities are initially small, they rapidly increase considerably. At the boundary, the electrons can reach thermal velocities of $6\ v_0$, much higher than the ion velocities. Because the electron and ion density profiles are very similar during the entire evolution of the system, the current density $J_y = e(n_iv_{iy}-n_ev_{ey})$ is then mainly transported by the electrons, where $n_j$ is the density and $v_{jy}$ the $y$ component of the velocity of the ions and electrons ($j=i,e$, respectively). This is also consistent with the observed spatial distribution of electrons during the reflection, which shows an excess of fast electrons around the standoff position.

\subsection{Driver length} \label{sec:length}

\par
\begin{figure*}[t]
    \includegraphics[width=\linewidth]{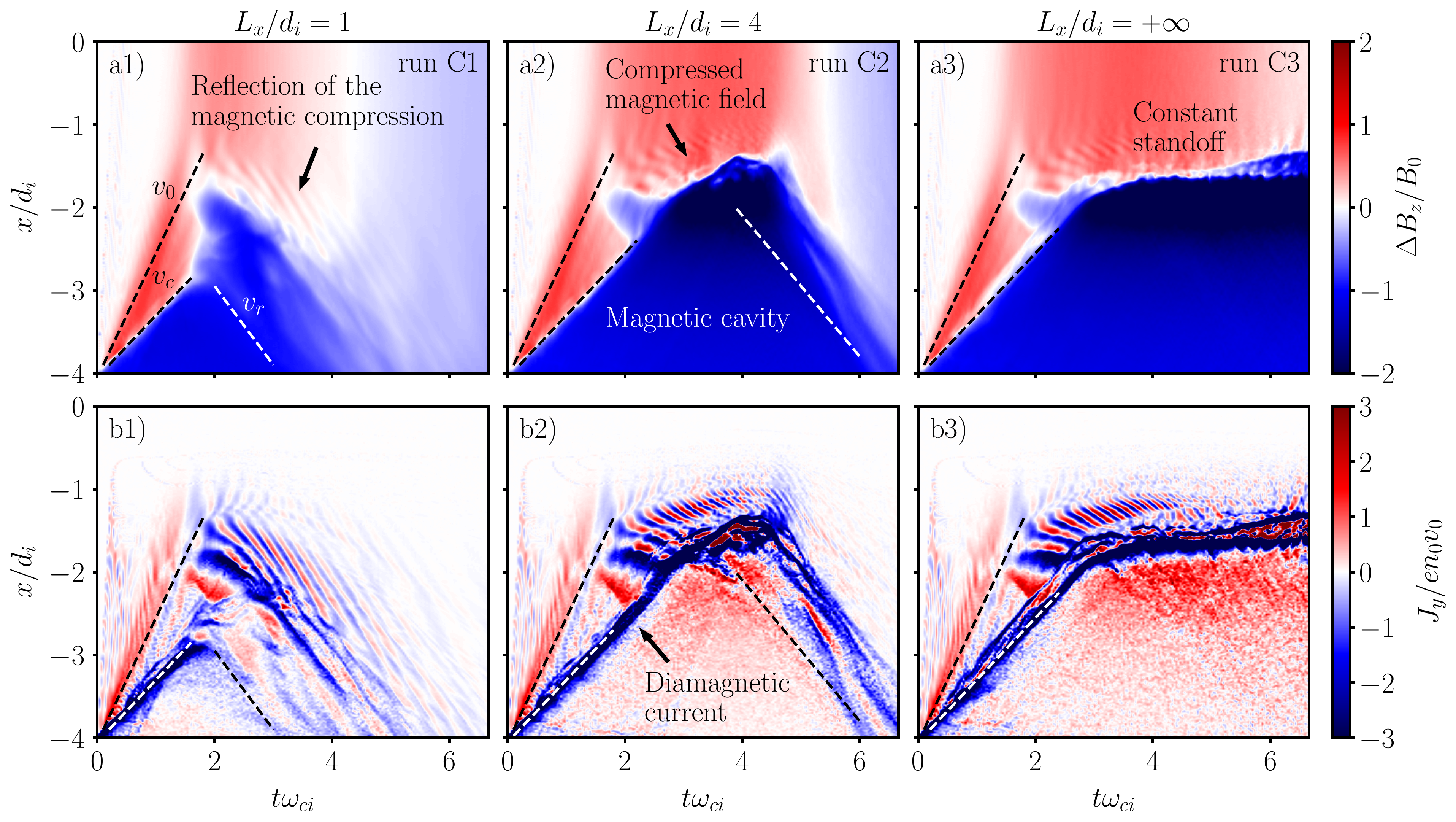}
    \caption{\label{fig:scan-length} Temporal evolution of the variations of the magnetic field $\Delta B_z$ and current density $J_y$ at $y=0$, for driver lengths of a) 1 $d_i$, b) 4 $d_i$ and for c) an infinite driver length (see Table~\ref{tab:runs} for a full list of the parameters). The dashed lines represent the slopes of the flow velocity $v_0$, the coupling velocity $v_c$, and the reflection velocity $v_r$.}%
\end{figure*}

 To choose a driver length that best reproduces the experimental results shown in Fig.~\ref{fig:experiment} and to understand its role on the magnetic field and current density structures, we performed simulations \textrm{C1} to \textrm{C3} (see Table~\ref{tab:runs}) with varying driver length $L_x$. In Fig.~\ref{fig:scan-length}, we show $\Delta B_z$ and $J_y$ at $y=0$ for $L_x=1\ d_i$ (C1), $L_x=4\ d_i$ (C2) and for an infinite driver (C3). For these simulations, the properties of the background plasma and the width of the driver $L_y$ were kept unchanged. The density of the driver was $n_d=2\ n_0$.%

\par In Figs.~\ref{fig:scan-length} a1) and b1), we see the magnetic field and current density plots for the short driver length $L_x=1\ d_i$. We observe most of the features of Fig.~\ref{fig:standard}, namely the reflection of the compressed magnetic field in a1) and the diamagnetic and magnetopause currents in b1). For this length, however, the driver never fully interacts with the dipole. The closest that the diamagnetic current structure gets to the dipole is $x_r\approx-3.0\ d_i$, \textit{i.e.}, much farther than the expected standoff $x_0 = -L_0=-1.8\ d_i$. To replicate the experimental results and ensure that the driver can reach the dipole, we should thus use a sufficiently long driver such that $x_r>x_0$. Additionally, short drivers risk entering in a decoupling regime between the two plasmas~\cite{Hewett2011}, which can compromise the observation of a magnetopause. The coupling effects on the results are discussed in detail in Sec.~\ref{sec:density}.%


\par The position where the driver is fully reflected by the background can be estimated as $x_r\approx x_B+L_xv_c/(v_0-v_c)$, where $x_B$ is the initial boundary position between the two plasmas. This estimate is obtained by computing the volume of the background plasma required for the driver plasma to deposit its kinetic energy, \textit{i.e.} $x_r - x_B$ corresponds to the magnetic stopping radius of the system~\cite{rb}.


\par In the simulation with $L_x=4\ d_i$, represented in Figs.~\ref{fig:scan-length} a2) and b2), we observe once more the main features identified in Fig.~\ref{fig:standard}, but unlike the $L_x=1\ d_i$ case, the driver is long enough and ends up reflected by the dipole. We observe that the diamagnetic current reaches the expected standoff and has enough plasma to maintain it near the dipole for a time period ($t\omega_{ci}\approx3$ to $t\omega_{ci}\approx5$) longer than the $2\ d_i$ case shown in Fig.~\ref{fig:standard}. As a result, the magnetic decompression in the background region is delayed for longer drivers. However, because the full driver reflection also occurs later, longer drivers will result in short-lived reflections of the compression of the magnetic field.%

\par In Figs.~\ref{fig:scan-length} a3) and b3), we show the results for a driver with infinite length ($L_x=+\infty$). In this simulation, the driver plasma is only partially initialized inside the simulation domain, and a flow is continuously injected from the lower $x$ boundary. An infinite driver configuration allows us to understand the dynamics of the system in an asymptotic regime in which the driver plasma stays close to the dipole. As expected, until $t\omega_{ci}=3$, the features observed are very similar to $L_x=2\ d_i$ and $L_x=4\ d_i$. After this time, the magnetic and the driver kinetic pressures balance each other near $x_0$, so the diamagnetic current remains stationary. Because the driver can hold for longer near the dipole, the decompression in the background region is much slower and is not visible for the time range of the plot. We can also observe that the background waves are only visible during a transient.%

\par In all the three simulations, the coupling velocity measured was always $v_c \approx 0.49\ v_0$. Given the results shown in Fig.~\ref{fig:scan-length}, we chose a driver length of $2\ d_i$ to reproduce the experimental results. This driven length is large enough to ensure that the driver arrives at the dipole and small enough to observe a significant reflection of the compression of the magnetic field as we see in the experiments.%

\subsection{Plasma coupling with density ratio} \label{sec:density}

\par
\begin{figure*}[t]
    \includegraphics[width=\linewidth]{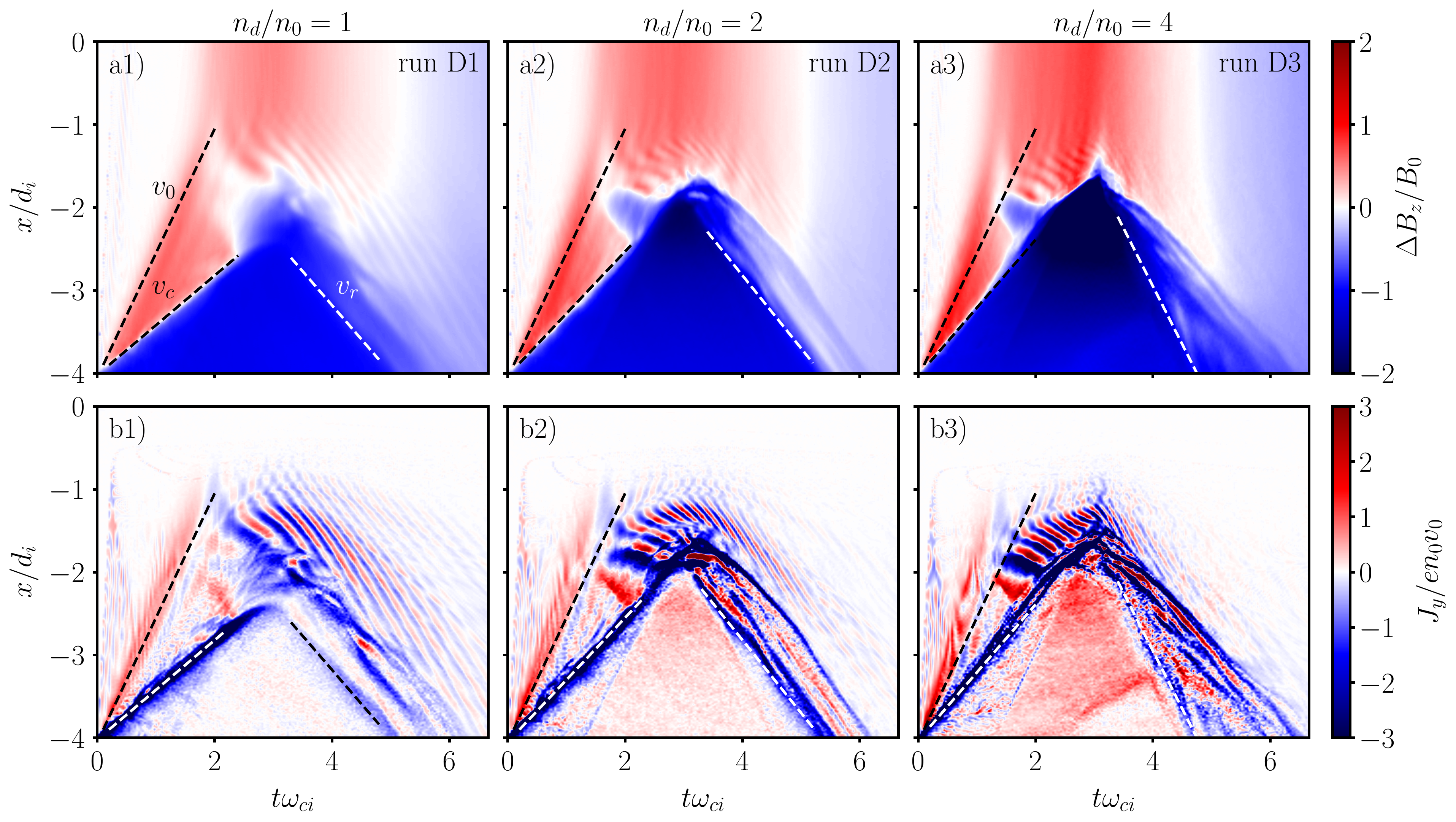}
    \caption{\label{fig:scan-density} Temporal evolution of the variations of the magnetic field $\Delta B_z$ and current density $J_y$ at $y=0$, for different ratios between the driver and background densities $n_d / n_0$. The magnetic moment was chosen so that the expected standoff distance $L_0$, calculated from Eq.~\eqref{eq:pressure-equilibrium}, was kept as 1.8 $d_i$ for all the simulations. Panels a-c) show results for $n_d = n_0$,  $n_d = 2\ n_0$ and $n_d=4\ n_0$, respectively.}%
\end{figure*}

 As expected from previous works, increasing the ratio between the driver and background plasma densities should improve the coupling between the two plasmas~\cite{Bondarenko2017, Hewett2011}, meaning that, for denser drivers, the transfer of momentum and energy from the driver to the background plasma is more efficient. To better understand the role of the coupling mechanism, we performed simulations with different values of the driver density, namely $n_d=n_0$ (\textrm{D1}), $n_d=2\ n_0$ (\textrm{D2}) and $n_d=4\ n_0$ (\textrm{D3}), while keeping a constant background density $n_0$ and a driver length $L_x = 2\ d_i$. For each run, the magnetic moment was chosen such that the expected standoff obtained from Eq.~\eqref{eq:pressure-equilibrium} was always $L_0=1.8\ d_i$. The synthetic magnetic field and current density diagnostics were obtained for these simulations and are shown in Fig.~\ref{fig:scan-density}.%

\par In Figs.~\ref{fig:scan-density} a1) and b1) we can see $\Delta B_z$ and $J_y$ for the lowest driver density considered, $n_d = n_0$ (\textit{i.e.}, background and driver with the same initial density). In this regime, the coupling is less efficient and, as a result, the coupling velocity $v_c \approx 0.38\ v_0$ is lower than obtained in the higher densities cases represented in Figs.~\ref{fig:scan-density} b) and c). Due to the low coupling velocity, the driver plasma is reflected more quickly by the background than for denser drivers, and the expected position $x_r$ for the total reflection on the background is farther from the dipole than the expected standoff $x_0$, meaning $x_r<x_0$. As a result, Fig.~\ref{fig:scan-density} a) shows similarities with the short driver length represented in Fig.~\ref{fig:scan-length} a), because, in both simulations, the driver parameters do not ensure that the driver arrives near the dipole.%

\par In Figs.~\ref{fig:scan-density} a2) and b2), we show the results for $n_d = 2\ n_0$, which is the same run represented in Fig.~\ref{fig:standard}. For this density, the coupling velocity, measured as $v_c\approx0.49\ v_0$, is high enough to secure a reflection of the driver by the dipole, as we observe at $t\omega_{ci}\approx3$. In Figs.~\ref{fig:scan-density} a3) and b3) we show the case with the highest driver density $n_d=4\ n_0$, which is similar to the $n_d=2\ n_0$ case, because the measured coupling velocity for Fig.~\ref{fig:scan-density} c) was $v_c \approx 0.56\ v_0$, \textit{i.e.}, only slightly larger than the $v_c$ measured for Fig.~\ref{fig:scan-density} b). In the high density case, we also see that the current density structures during the plasma reflection are filamented, due to analogous multi-stream velocity distributions discussed for Fig.~\ref{fig:scan-length} b).%

\par To guarantee that the driver reaches the expected standoff, we thus require that $x_r>x_0$. In fact, the position where the driver is reflected $x_r$, for no dipole cases, increases with the driver length $L_x$ and the velocity ratio $v_c/v_0$, and thus, both quantities must be large enough to guarantee that $x_r>x_0$. In turn, the ratio $v_c/v_0$ increases with increasing driver density ratio $n_d/n_0$, and so, the driver should be sufficiently long and dense to effectively couple to the background plasma. Our results (in particular Sec.~\ref{sec:additional}) show that a driver with $L_x=2\ d_i$ and $n_d=2\ n_0$ qualitatively reproduces the experimental results.%

\par A separate study was also performed to analytically determine the properties of the driver-background plasma coupling. The results of this study will be presented in a future paper.

\subsection{Dependency of the magnetopause position with the magnetic moment} \label{sec:momentum}

\par
\begin{figure*}[t]
    \includegraphics[width=\linewidth]{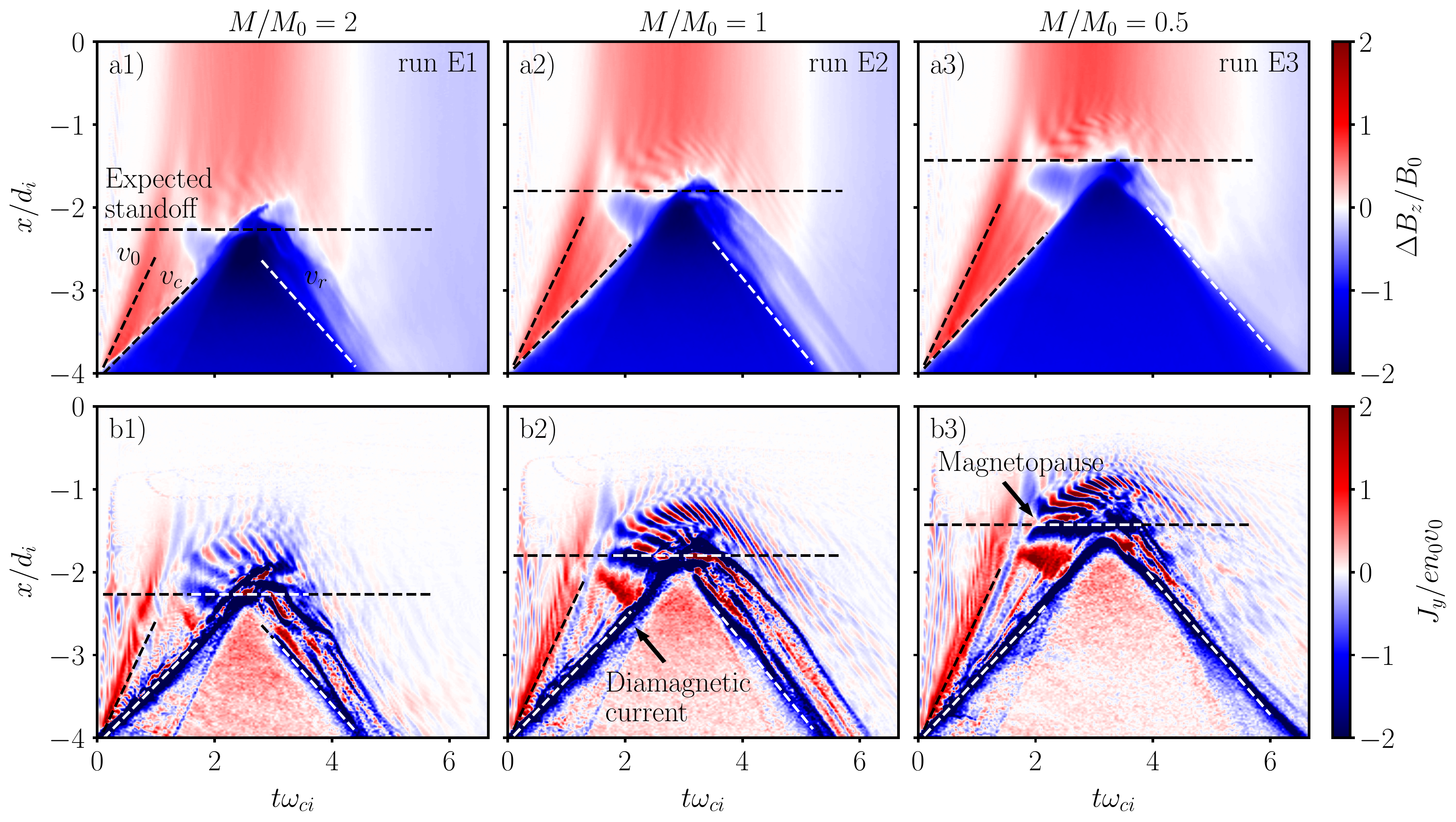}
    \caption{\label{fig:scan-momentum} Temporal evolution of the variation of the magnetic field $\Delta B_z$ and current density $J_y$ at $y=0$, for three different magnetic moments. The magnetic moments $M$ considered were a) $M=2\ M_0$, b) $M=M_0$ and c) $M=M_0/2$, where $M_0$ represents the magnetic moment that corresponds to a standoff $L_0=1.8\ d_i$ for a driver density $n_d=2\ n_0$. The corresponding standoffs for these simulations are a) $L_0\approx2.3\ d_i$, b) $L_0 = 1.8\ d_i$ and c) $L_0 \approx 1.4 \ d_i$.}%
\end{figure*}

 To confirm that the features previously associated with the magnetopause location change according with its expected position, we performed simulations with a 2 $d_i$ long driver with density $n_d = 2\ n_0$ for three different magnetic moments. Considering the magnetic moment that results in the expected standoff $L_0=1.8\ d_i$ as $M_0$ (simulation \textrm{B/E2} on Table~\ref{tab:runs}), simulations with the magnetic moments $2\ M_0$ (\textrm{E1}) and $M_0/2$ (\textrm{E3}) were also performed, corresponding respectively to the expected standoffs $L_0\approx2.3\ d_i$ and $L_0\approx1.4\ d_i$. Fig.~\ref{fig:scan-momentum} shows the $\Delta B_z$ and $J_y$ synthetic diagnostics at $y=0$ for the three simulations.%

\par Figs.~\ref{fig:scan-momentum} a1) and b1) show the results for the highest magnetic moment $M=2\ M_0$. We see that the current structures associated with the magnetopause and the background waves are less evident than for the lower magnetic moments, as they are formed farther from the dipole. Figs.~\ref{fig:scan-momentum} a2) and b2) correspond to the magnetic moment $M_0$ that leads to $L_0=1.8\ d_i$ and are the same results shown in Fig.~\ref{fig:standard}. As previously mentioned, there are two main observable current structure standoffs. The first one is associated to the diamagnetic current, which is reflected around $t\omega_{ci}\approx3$ near the expected value $x_0=-L_0=-1.8\ d_i$. This standoff is related to the interaction between the driver ions and the dipole. The second standoff occurs between $t\omega_{ci}\approx2$ and $t\omega_{ci}\approx3$ and it is located in the background plasma region. This standoff also occurs near $x=-1.8\ d_i$.%

\par In Figs.~\ref{fig:scan-momentum} a3) and b3), we show the results obtained for the half magnetic moment $M=M_0/2$. In this case, the magnetic pressure exerted by the dipole is lower, leading to a smaller $L_0$, and consequently, the diamagnetic current feature visible in b3) is closer to the dipole than in Figs.~\ref{fig:scan-momentum} b1) and b2). The main changes, however, occur in the magnetopause current. Unlike what we observe for the other magnetic moments, the magnetopause current, pinpointed in the current density plot, lasts for a longer time (until $t\omega_{ci}\approx4$). This current is also more separated from the diamagnetic current standoff and is easier to identify. This is consistent with the experimental observations.%

\par To identify the pressure balances associated with the two observed standoffs, and because the magnetic and kinetic pressures vary over time, we studied the temporal evolution of the different plasma and magnetic pressure components of the system. In particular, we calculated the spatial profiles of the magnetic pressure $B^2/8\pi$, the ram pressure $n_jm_jv_{flj}^2$ and the thermal pressure $n_jm_jv_{thj}^2$ as a function of time for $y = 0$. In these expressions, $n_j$, $m_j$, $v_{flj}$ and $v_{thj}$ refer to the density, mass and flow and thermal velocities, respectively, of the ions ($j=i$) and electrons ($j=e$). The magnetic pressure was calculated from the magnetic field measured in each PIC grid cell located at $y=0$. The flow and thermal pressures, were calculated from averaged particle data. To ensure that the calculation of each kinetic pressure considered a sufficiently large number of particles, all the particles between $-0.1\ d_i<y<0.1\ d_i$ were binned into equal-sized bins of width $0.05\ d_i$ over the $x$ direction. For each bin we computed: i) the average density of each species, ii) the flow velocity, corresponding to the average of the velocity of the particles, and iii) the thermal velocity, corresponding to the standard deviation of the velocity of the particles~\cite{Liboff2003}. With these averaged quantities, the ram and thermal pressures were calculated in each bin for each species of ions and electrons and each component of the velocity $x$, $y$, and $z$. The $y$ and $z$ components of the pressures, however, are negligible.

\par
\begin{figure*}[t]
\includegraphics[width=\linewidth]{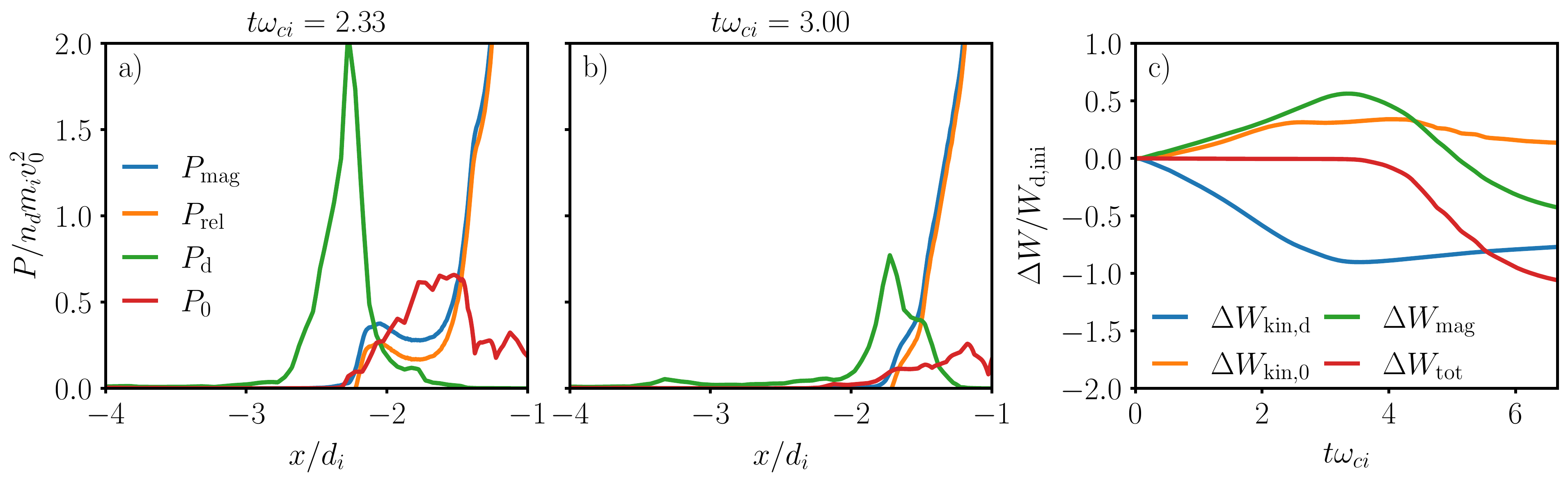}
\caption{\label{fig:pressures} Pressures profiles calculated for simulation \textrm{E3} with a magnetic moment $M=M_0/2$ (shown in Fig.~\ref{fig:scan-momentum} c)), during the occurrence of a) the magnetopause and b) the standoff of the diamagnetic current. The magnetic pressures are $P_\mathrm{mag} = B_z^2/8\pi$ and $P_\mathrm{rel} = P_\mathrm{mag}-B_0^2/8\pi$. The kinetic pressures $P_\mathrm{d}$ and $P_0$, corresponding to the driver and background plasmas, respectively, consider both the ions and electrons and the flow and thermal components of the velocity. c) Temporal evolution of the variation of the total kinetic energies of the driver $\Delta W_{\mathrm{kin,d}}$ and background $\Delta W_{\mathrm{kin,0}}$ plasmas, the magnetic energy $\Delta W_{\mathrm{mag}}$, and the total energy of the simulation box $\Delta W_{\mathrm{tot}}$. The total energy is calculated by adding all the kinetic energies and the electric and magnetic energies. Since the background plasma is magnetized, the electric energy term is many orders of magnitude smaller than the magnetic energy term. The energies were normalized to the initial total energy of the driver ions $W_{\textrm{d,ini}}$. The loss of energy conservation near $t\omega_{ci}\approx 4$ is caused by the escape of background plasma particles and magnetic field through the right hand side of the simulation box.}%
\end{figure*}

 These pressure profiles were obtained for simulation E3 with magnetic moment $M=M_0/2$ and are plotted in Fig.~\ref{fig:pressures} for times where a) the magnetopause and b) the diamagnetic current standoff can be observed. The kinetic pressures represented were calculated by adding all the components of the ram and thermal pressures of the ions and electrons for the background ($P_0$) and the driver ($P_\mathrm{d}$) plasmas. The magnetic pressures represented were calculated by considering the total and the relative magnetic field pressures ($P_\mathrm{mag} = B_z^2/8\pi$ and $P_\mathrm{rel} = P_\mathrm{mag}-B_0^2/8\pi$, respectively). The pressures were normalized to the initial ram pressure of the driver ions.%

\par Fig.~\ref{fig:pressures} a) shows the magnetic and kinetic pressures at time $t\omega_{ci}\approx2.33$ where we observed the magnetopause in Fig.~\ref{fig:scan-momentum} b3). When the driver starts pushing the background, the pressure of the driver at the interface between the plasmas increases because the driver density and thermal velocities also increase. During the flow, the driver transfers energy and momentum to the background plasma, and as a result, the background develops a strong kinetic pressure. At the time represented in Fig.~\ref{fig:pressures} a), the background plasma pressure equals the total and the dipolar magnetic pressures in $x \approx -1.4\ d_i$, near the location of the magnetopause current of Fig.~\ref{fig:scan-momentum} b3). This observation supports the hypothesis that this current emerges from the standoff between the background and magnetic pressures. Fig.~\ref{fig:pressures} b) shows the pressures for $t\omega_{ci}=3$ where we see the beginning of the reflection of the driver. The driver pressure equals the magnetic and dipolar pressures near $x_0=-L_0\approx-1.4\ d_i$. After this time, the driver is incapable of moving any further into the background because the magnetic pressure exceeds its kinetic pressure.%

\par The energy variations integrated over the entire simulation domain can also help us understand the system dynamics. Fig.~\ref{fig:pressures} c) shows the variation of the total driver and background kinetic energies, $\Delta W_{\mathrm{kin,d}}$ and $\Delta W_{\mathrm{kin,0}}$, respectively, as well as the variation of the magnetic energy $\Delta W_{\mathrm{mag}}$, and of the total energy $\Delta W_{\mathrm{tot}}$. The kinetic energies of the background and driver plasmas consider all ions and electrons. In early times $t\omega_{ci}<3$, as the driver and background plasmas interact, the driver transfers its energy to the background plasma and the magnetic field. The total energy, given by the sum of the electromagnetic energy and the kinetic energies, remains constant during this period. After the driver is fully reflected by the dipole for $t\omega_{ci}>3$, the magnetic field loses most of its energy to the background and driver plasmas leading to a drop of the magnetic energy. After $t\omega_{ci}\approx4$, the background ions start to leave the simulation box, and the total energy is no longer conserved. The background kinetic energy remains approximately constant because the background plasma loses energy to the sink at the right boundary of the simulation but gains energy from the magnetic field. For both driver and background plasma, the ions carry most of the energy.%

\par
\begin{figure}[ht]
\includegraphics[width=\columnwidth]{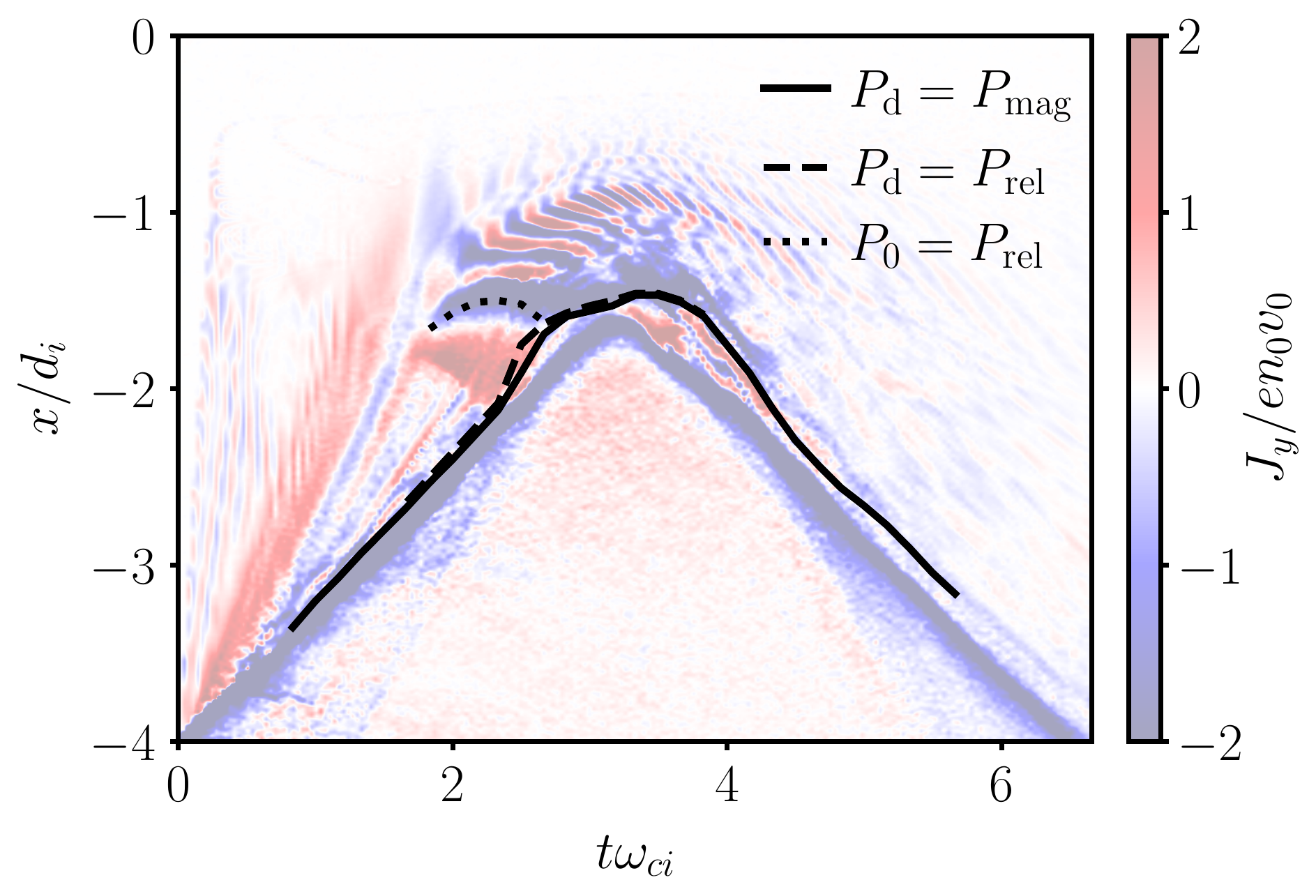}
\caption{\label{fig:equilibrium} Temporal evolution of the current density $J_y$ at $y=0$, with the closest locations to the dipole of different pressure balances for multiple times. The represented locations of pressure balances are the equilibria between the driver kinetic pressure $P_{\mathrm{d}}$ with the total magnetic field pressure $P_{\mathrm{mag}}=B_z^2/8\pi$, represented by the solid line; the background kinetic pressure $P_0$ with the pressure exerted by the relative magnetic field $P_{\mathrm{rel}}=P_{\mathrm{mag}}-B_0^2/8\pi$, by the dotted line, and $P_{\mathrm{d}}=P_{\mathrm{rel}}$, by the dashed line. The results correspond to simulation \textrm{E3} (see Table~\ref{tab:runs}).}%
\end{figure}

 From Fig.~\ref{fig:pressures}, we can identify the positions where multiple pressure balances occur, and therefore, develop an insight into the pressure equilibria that are behind the structures of the current density synthetic diagnostics. Using the previously calculated pressures, we obtained the equilibrium positions where certain pressure balances manifested and plotted them in Fig.~\ref{fig:equilibrium} alongside $J_y$.%

\par This analysis shows that the system has, in general, two magnetopause structures: one driven by the background, and one by the driver plasma. The former structure is defined by the balance $P_{\mathrm{0}}=P_{\mathrm{rel}}$. For the latter structure to form, the driver needs to have almost enough energy to push the diamagnetic current up to the magnetopause, defined by Eq.~\ref{eq:pressure-equilibrium}. This is illustrated in Fig.~\ref{fig:equilibrium}, where we show the location of the pressure equilibrium between the driver kinetic pressure and the total magnetic pressure, $P_\mathrm{d} = P_\mathrm{mag}$. 

\par As shown in Fig.~\ref{fig:pressures}, the current associated with the background magnetopause seems to overlap with the region of background and magnetic pressure balance. Unlike the driver, the background plasma is magnetized. If we neglect the compression of the magnetic field in the downstream region, the pressure balance that describes this magnetopause can then be estimated by the equilibrium of the kinetic pressure of the background plasma with the relative magnetic pressure, $P_0 = P_\mathrm{rel}$. In Fig.~\ref{fig:equilibrium}, we show that this pressure balance, represented by the dotted line, describes well the position of the current feature identified as the magnetopause between times $t\omega_{ci}\approx2$ and $t\omega_{ci}\approx3$. 


\par After $t\omega_{ci} \approx 3$, the magnetopause current is well described by the pressure balance $P_\mathrm{d} = P_\mathrm{rel}$, as illustrated by the dashed line in Fig.~\ref{fig:equilibrium}. In fact, after inspecting the phase spaces in Figs.~\ref{fig:phase-space} a3) and b3), we can observe that a combination of driver plasma particles (separated from the bulk distribution) and background ions pushes the dipolar field and sets the position of the magnetopause.

\par We stress that, because we are determining equilibria via MHD pressure balances but are checking the intersection between pressure curves with kinetic resolution, some caution must be made to ensure that we are observing the equilibrium between pressures and not merely the interface between the different regions of interest. To ensure that the pressure equilibria were correctly obtained, the corresponding pressure profiles were always carefully inspected with additional diagnostics.%

\subsection{Realistic parameters}\label{sec:realistic}

\par Due to the need for more extensive scans (and thus using physically equivalent but computationally feasible parameters), the simulations shown so far considered reduced ion mass ratios, cold plasmas, and higher velocities than the ones used in the LAPD experiments - see Table~\ref{tab:parameters}. To ensure that the main results presented in the previous sections are also valid with realistic parameters, we have performed a set of simulations with parameters similar to those expected experimentally. 

\par Three simulations were performed, labeled as runs F1 to F3. Run F1 employs realistic mass ratios $m_{i,d}/{m_e} = m_{i,0}/{m_e} = 1836$. Additionally, run F2 also considers a ratio between the electron thermal and flow velocities close to the ones expected for the LAPD experiments, namely $v_{the,x}/v_0=2.5$ and $v_{thi,x}/v_0=0.033$, leading to higher temperatures than in the previous simulations, and thus allowing possible thermal effects on the system. Finally, run F3 considers the same electron thermal velocity ratios of F2 but the standard reduced mass ratios. 

\par
\begin{figure*}[t]
    \includegraphics[width=\linewidth]{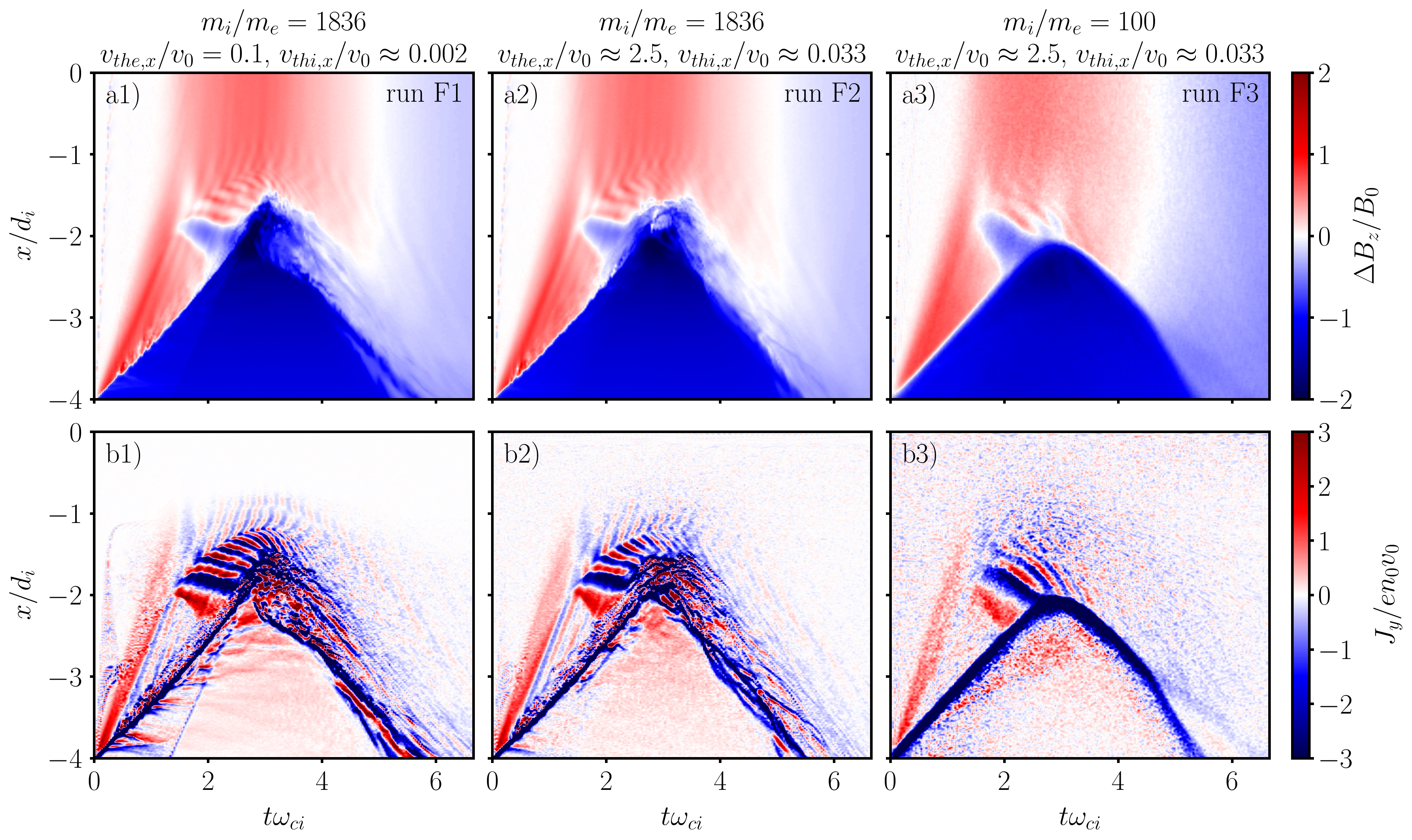}
    \caption{\label{fig:realistic} Temporal evolution of a) the variation of the magnetic field $\Delta B_z$ and b) the current density $J_y$ at $y=0$, for the simulations with similar parameters to the experiments. Run F1 considers realistic mass ratios for the driver and background plasmas and low ratios between the thermal and flow velocities; run F2 uses realistic mass ratios and thermal velocity ratios close to the ones expected in the experiments; run F3 uses the realistic thermal velocity ratios but reduced mass ratios.}%
\end{figure*}

 The $\Delta B_z$ and $J_y$ plots for these simulations are shown in Fig.~\ref{fig:realistic}. Note that, due to changes in $m_i/m_e$, the spatial and temporal scales were recalculated for the new parameters. Once again, the magnetic dipole moment for the three simulations was adjusted to ensure that $L_0=1.8\ d_i$.%

\par As expected, these simulations show the same main structures discussed in the previous sections. We observe the typical reflection of the compression of the magnetic field and the current structures of the magnetopause and diamagnetic cavity. However, some differences are also visible. In Figs.~\ref{fig:realistic} a1) and b1), \textit{i.e.} for the realistic mass ratios but cold plasmas simulation, we observe a stronger filamentation of the plasma flow reflected off the dipole and a thinner diamagnetic current. This is because $d_e$ is the characteristic length scale of the current layer and we have lower $d_e/d_i$ values for larger $m_i/m_e$. Figs.~\ref{fig:realistic} a2) and b2), for the simulation with higher temperatures, show no major differences with Figs.~\ref{fig:realistic} a1) and b1), even though there is a significant increase in the thermal velocities. 

\par In Figs.~\ref{fig:realistic} a3) and b3), however, we observe significant differences for reduced mass ratios with realistic thermal velocity ratios. In particular, we observe in the current density plot smoother magnetic and current structures and less defined background waves between the magnetopause and the dipole. We also observed for increased ion thermal velocities, for example, $v_{thi}/v_0 \approx 0.25$, that the background waves are no longer visible. 

\par Additionally, other simulations were performed to look for possible changes with realistic parameters. A simulation with a lower flow velocity $v_0=0.01\ c$ and realistic thermal velocity ratios lead to no significant features observed, and the obtained synthetic diagnostics were very similar to the ones in Figs.~\ref{fig:realistic} a3) and b3), meaning that the system scales well with $v_0$. Another simulation was performed to observe if the shape of the initial density profiles of the plasmas would affect the main results. Namely, the constant density profiles used on both the driver and background plasmas were replaced by Gaussian density profiles with a typical gradient scale $\sigma=1\ d_i$ on the edges of the plasmas. This simulation did not show meaningful differences, in agreement with previous plasma coupling works, which observed that the leading edge of the plasmas evolves similarly for different initial density profiles~\cite{doi:10.1063/1.1694472}.%

\subsection{Finite transverse size}\label{sec:finite}

\par
\begin{figure*}[t]
    \includegraphics[width=\linewidth]{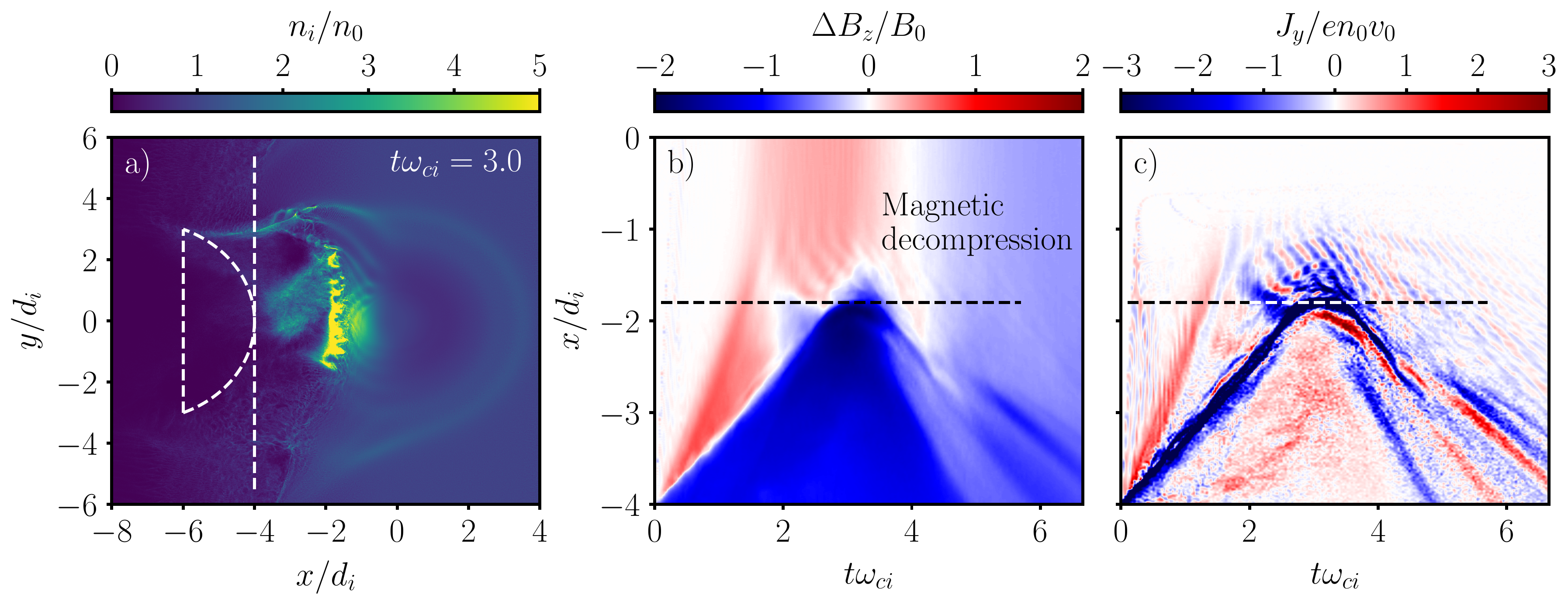}
    \caption{\label{fig:finite} a) Total ion density at time $t\omega_{ci}=3.0$, and temporal evolution of b) the variation of the magnetic field $\Delta B_z$ and c) the current density $J_y$ at $y=0$, for simulation G with a finite width driver with a circular segment shape. The dashed lines at a) represent the initial position of the driver and the left border of the background plasma.}%
\end{figure*}

 For simplicity, and because we were more interested in studying the system along the axis of symmetry $y=0$, the previous simulations only considered a driver with infinite width $L_y$ and a length of $L_x=2\ d_i$. In the experiments, however, the drivers had a width comparable to their lengths and did not have the sharp boundaries used in the simulations. To investigate if and how our results are modified with a more complex-shaped driver, we performed a simulation with a finite width, semi-circular-shaped driver plasma. This driver is initially defined with the conditions $(x+7.25\ d_i)^2+y^2<(3.25\ d_i)^2$ and $x>-6\ d_i$ and has length $L_x=2\ d_i$ and width $L_y=6\ d_i$. Fig.~\ref{fig:finite} shows the results of this simulation and includes the initial shape of the driver in Fig.~\ref{fig:finite} a).%

\par Due to the finite width of the new driver and its particular shape, we should expect to see significant differences in the regions of the simulation plane far from $y=0$. In the total ion density plot of Fig.~\ref{fig:finite} a) for a time $t\omega_{ci}=3$, when there is a strong interaction of the driver with the dipolar magnetic field, we observe the propagation of waves at the lower and upper sides of the dipole caused by the finite width of the driver, that was not present for infinite width drivers.%

\par In Figs.~\ref{fig:finite} b) and c), we see the usual magnetic and current density plots at $y=0$ for this simulation. By shortening the driver plasma width, the background particles escape from the bottom and top regions of the simulation box, and the driver has more difficulty holding the magnetic decompression in the background region. The decompression, therefore, occurs quicker for finite drivers, as seen in Fig.~\ref{fig:finite} b), leading to short reflections of the magnetic compression. 

\par Although this complex-shaped driver gets us closer to the experimental configuration, the simulations did not include all the properties of the experimental driver, as for example, the non-uniform density, velocity profiles of the plasmas and the flow divergence. Additionally, 3D effects should also be considered. Future simulations are planned to study the effect of these properties in the results. However, we expect that these features will not change the main results of the simulations.%

\section{Conclusions} 
\label{sec:conclusions}

\par In this work, we have performed PIC simulations of mini-magnetospheres in the interaction between a plasma flow and a magnetized background plasma. In particular, we have successfully reproduced results from recent experiments performed at the LAPD, validating the experimental platform to study mini-magnetospheres in the laboratory. We have also explored an extensive parameter space defining the interaction, allowing us to i) determine how the main properties of the system change with the parameters and ii) identify the required conditions for the creation of a mini-magnetosphere.%

\par Our simulations have shown that some system features are present across multiple regimes. The initial flow of the driver expels the magnetic field in the upstream region, leading to a magnetic cavity, and compresses the downstream magnetic field. The driver travels through the background until the magnetic field pressure is large enough to counterbalance the driver plasma pressure. A fast decompression of the background magnetic field then follows. If the background decompression occurs after the total reflection of the driver plasma, then we can observe the reflection of the compression of the magnetic field. To see this feature, the driver needs to be short enough to anticipate the driver reflection relative to the decompression but sufficiently long to ensure that it can get close to the dipole.%

\par For the super-Alfvénic flows considered, the driver particles are reflected upstream during the interaction with the background plasma and the magnetic field. The coupling velocity (\textit{i.e.}, the velocity at which the leading end of the driver travels through the background) is lower than the flow velocity and increases with the increase of the ratio between the driver and background densities. The coupling velocity and the length of the driver determine how far the driver can go through the background region without a dipole, for a uniform driver plasma. 

\par The interaction of the plasmas with the dipole results in two magnetopauses. The first describes the balance between the kinetic pressure of the propelled background plasma plus the pressure of the plasma internal magnetic field and the total magnetic pressure. The seconds describes approximately the balance between the kinetic pressure of the driver plasma separated from the bulk distribution and the relative magnetic pressure. Using simulations with different dipole moments, we have shown that, for lower magnetic moments, the driver and background standoffs are closer to the center of the dipole, and the magnetopause current is more clearly identified than for higher magnetic moments. Furthermore, it is also easier to separate the magnetopause and diamagnetic currents for lower magnetic moments, consistent with experimental observations.%

\par In the simulations performed, we also observed the formation of waves in the background plasma region, between the magnetopause and the center of the dipole, where the magnetic field gradient was significant. These waves result from the excitation that always followed the formation of the magnetopause and were only observed for background plasmas with relative low ion thermal velocities. This condition may explain the absence of these waves in the experimental plots.%

\par Most of the simulations presented in this work were performed in idealized configurations. In particular, we used reduced ion-to-electron mass ratios, unrealistically high flow velocities, a simple flat-top driver density profile, and neglected thermal effects. In Sec.~\ref{sec:realistic} and~\ref{sec:finite}, we presented simulations that drop some of these simplifications. Replacing reduced ion mass ratios with realistic ones and considering high thermal velocities ratios close to the obtained in the experiments did not lead to significant changes in the results. The same occurred when considering smoothed density profiles. It was also possible to conclude that the main features of the system scaled as expected with the absolute value of the driver flow velocity. We also presented a simulation to study possible effects associated with the complexity of the experimental laser-ablated driver. A simple circular segment-shaped driver was considered and led to similar results in the axis of symmetry as the infinite width driver simulations. However, wave-like structures were observed on both the bottom and upper sides of the dipole. For future studies on the regions outside the axis of symmetry, the driver shape and complexity must be considered.%

\par Additionally, we also performed other parameter scans related to the complexity of the driver. For instance, we performed simulations where the driver ions were heavier than the background ions to simulate the small role of the carbon ions in the experimental driver. These studies showed no significant differences to the lighter ions simulations.

\par In conclusion, the simulations were consistent with the LAPD experimental results, and the multiple parameter scans performed dictated the formation conditions of the main features of mini-magnetospheres. For future works, we intend to exploit the features present in the sides of the dipole, exploit anti-parallel magnetic field configurations, perform 3D simulations, and consider even more realistic properties of the driver.

\begin{acknowledgments}

\par We acknowledge the support of the European Research Council (InPairs ERC-2015-AdG 695088), FCT (PD/BD/114307/2016 and APPLAuSE PD/00505/2012), the NSF/DOE Partnership in Basic Plasma Science and Engineering (Award Number PHY-2010248), and PRACE for awarding access to MareNostrum (Barcelona Supercomputing Center, Spain). The simulations presented in this work were performed at the IST cluster (Lisbon, Portugal) and at MareNostrum.

\end{acknowledgments}

\nocite{*}

\bibliography{apssamp}

\end{document}